\documentclass[pdflatex,sn-mathphys-num]{sn-jnl}% Math and Physical Sciences Numbered Reference Style
\usepackage{silence}
\hypersetup{bookmarksdepth=2}
\usepackage{tikz}
\usepackage{graphicx}%
\usepackage{float}%
\usepackage{flafter}%
\usepackage{placeins}%
\usepackage{subcaption}%
\usepackage{multirow}%
\usepackage{amsmath,amssymb,amsfonts}%
\usepackage{amsthm}%
\usepackage{mathrsfs}%
\usepackage[title]{appendix}%
\usepackage{xcolor}%
\usepackage{textcomp}%
\usepackage{manyfoot}%
\usepackage{booktabs}%
\usepackage{algorithm}%
\usepackage{algorithmicx}%
\usepackage{algpseudocode}%
\usepackage{listings}%
\usepackage{caption}

\theoremstyle{thmstyleone}%
\theoremstyle{thmstyletwo}%
\theoremstyle{thmstylethree}%

\begin{document}

\title[Superconducting Qubits up to 2 K]{Characterisation of Superconducting Qubits up to 2 K}

\author*[1]{\fnm{Songyuan} \sur{Zhao}}\email{songyuan.zhao@glasgow.ac.uk}
\author[1,2]{\fnm{Valentino} \sur{Seferai}}
\author[1,2]{\fnm{Wridhdhisom} \sur{Karar}}
\author[1,2]{\fnm{Martin} \sur{Weides}}

\affil*[1]{\orgdiv{James Watt School of Engineering}, \orgname{University of Glasgow}, \orgaddress{\country{United Kingdom}}}
\affil[2]{\orgname{Quantcore Technologies Ltd.}, \orgaddress{\city{Glasgow}, \country{United Kingdom}}}

\abstract{Superconducting transmons are conventionally operated at millikelvin temperatures, motivating the development of Josephson-junction technologies with larger superconducting energy gaps. In this study, we fabricate Nb/Al/AlO$_x$/Al/Nb multilayer transmons and investigate their microwave response between 50 mK and 2 K. Usadel-equation calculations using literature-based material parameters predict a proximity-enhanced gap in a 10 nm Al layer of approximately 85\% of the bulk Nb value. At 50 mK, a device with an 8.755 GHz qubit transition exhibits energy-relaxation times of 1.5-2.5 µs and Ramsey dephasing times of 2.0-3.0 µs. Time-domain measurements track the evolution of these coherence times to 400 mK. Power-dependent resonator spectroscopy reveals an identifiable dispersive signature up to 2 K, albeit with substantial deterioration near the highest temperature. The dressed-resonance frequency shift and inverse quality-factor change exhibit Bose-Einstein-like temperature dependences, consistent with a driven open-system circuit-QED model. In contrast, an Al-based reference circuit measured in the same cryostat shows substantial degradation by 500 mK. These results establish the persistence of a coupled microwave response at elevated temperature.}


\maketitle

\section{Introduction}

The superconducting transmon qubit~\cite{Koch_2007_transmon} is an important architecture in the field of quantum computing and quantum information processing. It consists of a Josephson junction shunted by a large capacitance, which suppresses sensitivity to charge noise while retaining sufficient anharmonicity for selective microwave control~\cite{Kjaergaard_2020_review_qubits,Wendin_2017_review}. At present, transmons typically operate at $4$-$8\,\mathrm{GHz}$ and require dilution-refrigerator temperatures near $20$-$50\,\mathrm{mK}$ to suppress decoherence arising from thermally excited quasiparticles and thermal photons. Overcoming this conventional range of operating frequencies and temperatures is a key challenge for the scalability of next-generation superconducting quantum computing architectures~\cite{Wang_2026_all_nitride_ALD,Anferov_2024_20GHz_Qubits}.

In conventional transmon systems based on aluminium Josephson junctions~\cite{Koch_2007_transmon,Clarke_2008_superconducting_qubits,Oliver_2013_materials}, the critical temperature is approximately $1.2\,\mathrm{K}$, corresponding to a superconducting gap $\Delta\sim180\,\mu\mathrm{eV}$ and a pair-breaking frequency of $f_{\mathrm{pb}}=2\Delta/h\sim90\,\mathrm{GHz}$, where $h$ is the Planck constant. To suppress dissipative loss induced by quasiparticles, the circuit frequencies must satisfy $f_{\mathrm{r,q}}\ll f_{\mathrm{pb}}$~\cite{Mattis_1958_anomalous_skin}, where $f_{\mathrm{r}}$ and $f_{\mathrm{q}}$ are the readout-resonator and qubit transition frequencies, respectively. Furthermore, the balance between quantum and thermal effects in a microwave circuit is governed by the dimensionless ratio $hf/k_{\mathrm{B}}T$, where $hf$ is the photon energy at frequency $f$, $k_{\mathrm{B}}$ is Boltzmann's constant, and $T$ is the bath temperature of the circuit. In order for thermal-occupation effects to be small and the circuit system to operate in the quantum regime, $hf\gg k_{\mathrm{B}}T$ needs to be satisfied. Together, the superconducting energy scale and the requirement to suppress thermal occupation help motivate the conventional frequency and millikelvin operating ranges. 

Fault-tolerant quantum computation is likely to require thousands to millions of physical qubits. Achieving this scalability poses significant engineering challenges for existing transmon architectures~\cite{Devoret_2013_outlook, Fowler_2012_surface_codes, Arute_2019_quantum_supremacy}. One important constraint is the limited cooling power of dilution refrigerators at base temperatures of tens of millikelvin, which is typically only tens of microwatts. At the operating temperature of a given quantum-circuit technology, for example, the $20$-$50\,\mathrm{mK}$ range for Al qubits, the available cooling power imposes a critical limit on the total heat dissipated, including that generated by cabling, interconnects, and other passive and active microwave components. This thermal constraint fundamentally limits the number of qubits that can be operated, independently of subsequent implementation choices. At fixed $^{3}\mathrm{He}$ circulation, the mixing-chamber cooling power of a dilution refrigerator scales approximately as $T^2$~\cite{Wheatley_1968, deWaele_2011}. Thus, a tenfold increase in operating temperature could, in principle, provide approximately a hundred times greater cooling power, substantially relaxing a key cryogenic constraint on quantum processor scaling~\cite{Raicu_2025}.

Realising this cooling-power advantage requires circuit frequencies to increase with operating temperature such that $hf/k_{\mathrm{B}}T$ remains sufficiently large and thermal occupation remains suppressed. Increasing the superconducting gap is therefore a prerequisite for scaling both operating frequency and temperature without introducing excessive quasiparticle loss. At present, leading superconducting-qubit processors developed by Google, IBM, Rigetti, and IQM rely on Al-based Josephson junctions~\cite{Arute_2019_quantum_supremacy,Kim_2023_IBM_processor,Reagor_2018_Rigetti_processor,Abdurakhimov_2024_IQM_processor}. Recent wafer-scale technologies developed at VTT likewise retain Al-based junction materials, including Al/AlO$_x$/Nb trilayer structures~\cite{Sethi_2025_VTT_trilayer}. Increasing the superconducting energy scale beyond that of Al could be an important step toward translating the greater cooling power available at elevated temperatures into scalable quantum processors.

Recent research has sought to address this challenge by developing Josephson junctions based on materials with larger superconducting gaps. The two main approaches are junctions based on high-gap nitride materials~\cite{Nakamura_2011_epitaxial_qubits,Kim_2021_all_nitride_qubits,Wang_2026_all_nitride_ALD,Terai_2026_nitride_review} and niobium-aluminium multilayer junctions~\cite{Anferov_2024_20GHz_Qubits,Anferov_2024_Nb_Trilayer,Anferov_2025_mmWave_Qubits}, in which the superconducting properties of the multilayer are shaped by the proximity coupling between the superconducting layers to result in higher superconducting energy gaps~\cite{Zhao_1993_NbAl_proximity, Songyuan_2018, Zhao_2018, Usadel_1970_generalized}. Here, we focus on the latter approach, which retains the well-established amorphous AlO$_x$ tunnel barriers and their mature device physics~\cite{Anferov_2024_Nb_Trilayer,Zhao_1993_NbAl_proximity}.

The superconducting transition temperature $T_\mathrm{c}$ of niobium, typically in the range of $8$-$10\,\mathrm{K}$, is significantly higher than that of aluminium.  This higher $T_\mathrm{c}$, in turn, corresponds to a larger energy gap in the superconducting density of states. In Nb/Al/AlO$_x$/Al/Nb junctions, the outer niobium electrodes are proximity-coupled to the thin Al interlayers, resulting in larger superconducting gaps compared to bare aluminium junctions. Anferov \emph{et al.} demonstrated an optically defined niobium multilayer junction process that yielded coherence times up to $62\,\mu\mathrm{s}$ with reduced quasiparticle sensitivity above $160\,\mathrm{mK}$~\cite{Anferov_2024_Nb_Trilayer}. Based on this material platform, transmon qubits operating at $11$-$24\,\mathrm{GHz}$ were shown to maintain coherence above $200\,\mathrm{mK}$~\cite{Anferov_2024_20GHz_Qubits}, and a millimetre-wave qubit at $72\,\mathrm{GHz}$ was operated at $0.87\,\mathrm{K}$ using only $^4$He refrigeration~\cite{Anferov_2025_mmWave_Qubits}. These previous studies established Nb/Al/AlO$_x$/Al/Nb multilayer junctions as a promising platform for extending qubit operation to higher frequencies and temperatures, with coherent operation demonstrated at temperatures below $1\,\mathrm{K}$, reaching $0.87\,\mathrm{K}$~\cite{Anferov_2024_20GHz_Qubits,Anferov_2025_mmWave_Qubits}. 

In the present work, we extend the characterisation of the dispersively coupled resonator-qubit response to $2\,\mathrm{K}$, substantially beyond the temperature range of these earlier demonstrations. We further investigate the quantitative relationship between the Nb and Al layer thicknesses and the resulting proximity-modified energy gap. Alongside this material analysis, we examine the measured dispersive and dissipative responses in terms of the occupation-dependent Jaynes-Cummings Hamiltonian and thermal Lindblad dynamics, supported by numerical simulations of open quantum-system dynamics.

To investigate these material and coupled-system effects, we fabricated transmon qubits based on Nb/Al/AlO$_x$/Al/Nb Josephson junctions. For the nominal $120\,\mathrm{nm}$ Nb/$10\,\mathrm{nm}$ Al geometry, Usadel-equation calculations using literature-based material parameters and an assumed Nb/Al bilayer interface predict a proximity-enhanced Al-layer energy gap of approximately $85\%$ of the bulk Nb gap. We characterised the devices using spectroscopic and time-domain measurements at $50\,\mathrm{mK}$, and subsequently measured the temperature dependence of the dispersively coupled readout resonance from $50\,\mathrm{mK}$ to $2\,\mathrm{K}$. The results were compared with those obtained from an Al transmon qubit measured using the same cryogenic measurement system. 

The resonance frequencies and Q-factors of the dressed and decoupled resonances were extracted by performing circle fits at each temperature. Comparing the dressed and decoupled quantities allowed the intrinsic temperature dependence of the resonator to be separated from that arising from the resonator-qubit interaction. The dressed-resonance detuning and change in inverse Q-factor relative to their $50\,\mathrm{mK}$ values both follow a Bose-Einstein functional form. This dependence can be understood through the third-order occupation-dependent term in the Jaynes-Cummings Hamiltonian and the Bose-Einstein occupation factors entering the thermal Lindblad dynamics. QuTiP simulations incorporating these processes reproduce the observed Bose-Einstein temperature dependence of both quantities. Crucially, the separation between the low-power and high-power readout resonances remains identifiable up to $2\,\mathrm{K}$, providing evidence for the persistence of a coupled-circuit dispersive signature at elevated temperatures. Our results demonstrate that Nb/Al/AlO$_x$/Al/Nb multilayer transmons can sustain dispersive coupling at temperatures far exceeding those accessible to conventional Al-based devices, thereby positioning them as promising technologies for scalable quantum architectures.

\section{Multilayer Qubit Design and Fabrication}
\begin{figure}[ht]
    \centering
    \includegraphics[width=0.8\textwidth]{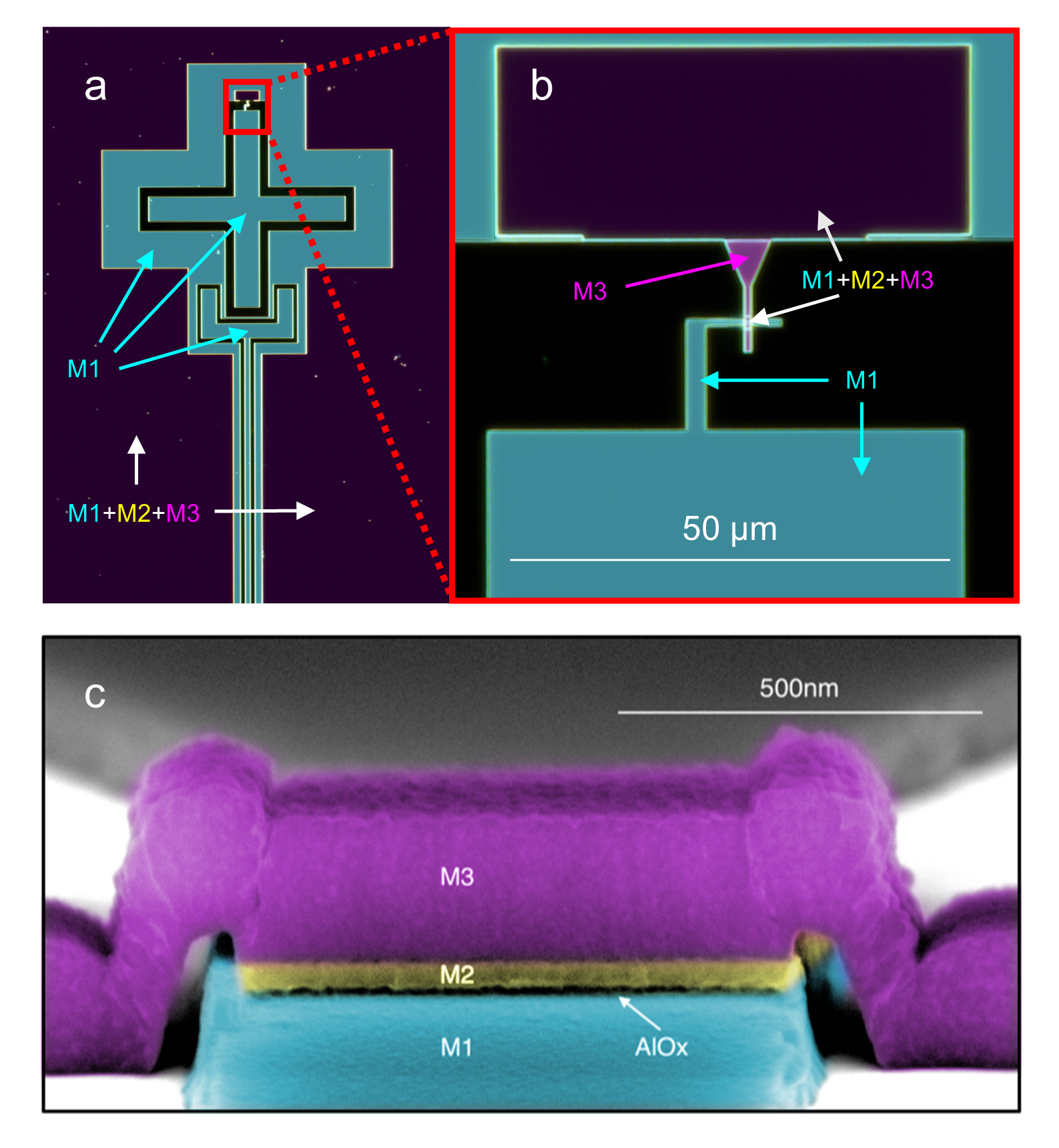}
    \caption{\textbf{Micrographs of the Nb/Al/AlO$_x$/Al/Nb qubit and Josephson junction.} \textbf{a}~Optical dark-field micrograph of the qubit island. The central cross spans $400\,\mu\mathrm{m}$ from end to end and has a width of $50\,\mu\mathrm{m}$. \textbf{b}~Zoomed micrograph around the multilayer Josephson junction. \textbf{c}~Scanning electron micrograph of the multilayer Josephson junction. The layers M1, M2, and M3 are niobium, with an Al/AlO$_x$/Al barrier stack deposited between M1 and M2. To better visualise the device metallisation, metal-covered regions are highlighted in cyan, yellow, and magenta for the M1, M2, and M3 layers, respectively.}%
    \label{fig:device_image}
\end{figure}

The multilayer Josephson junctions were fabricated at the James Watt Nanofabrication Centre using a fabrication process adapted from Anferov \emph{et al.}~\cite{Anferov_2024_Nb_Trilayer}. Sapphire substrates were cleaned in a piranha solution comprising sulphuric acid and hydrogen peroxide, rinsed in reverse-osmosis water, and dried with nitrogen. The Nb/Al/AlO$_x$/Al/Nb stack was deposited in situ, without breaking vacuum. The deposition sequence comprised first Nb base electrode deposition, followed by Al deposition, static oxidation at $2000\,\mathrm{Pa}$ for $60\,\min$, Al capping, and finally deposition of the Nb counterelectrode. The stack was patterned by electron-beam lithography in approximately $400\,\mathrm{nm}$ of electron-beam resist. An oxygen-plasma descum was performed immediately before the first dry etch, which used a $\mathrm{Cl}_2/\mathrm{BCl}_3/\mathrm{Ar}$ chemistry in an inductively coupled plasma reactive-ion etching (ICP-RIE) system to define the base electrode. The etch endpoint was monitored interferometrically, with a nominal $10\%$ over-etch. Subsequent dielectric isolation, wiring deposition, and selective plasma etching completed the junction and circuit definition, with the upper Nb wiring layer patterned using a $\mathrm{CF}_4/\mathrm{O}_2$ plasma, following the process sequence of Ref.~\cite{Anferov_2024_Nb_Trilayer}. Complete removal of the sidewall spacer was challenging because the required cleaning conditions could also attack the tunnel junction. The cleaning process was therefore limited to minimise junction damage, and residual dielectric may contribute to two-level-system (TLS) loss in the measured devices. Resist stripping, oxygen-plasma cleaning, and a final wet clean were then used to remove the remaining process material and surface oxide.

The device architecture is shown in Fig.~\ref{fig:device_image}. The optical dark-field micrographs in the upper panel show the qubit island where the central cross spans $400\,\mu\mathrm{m}$ from end to end and has a width of $50\,\mu\mathrm{m}$. From bottom to top, the multilayer junction comprises a $120\,\mathrm{nm}$ Nb base electrode (M1), followed by a $10\,\mathrm{nm}$ bottom Al layer, an AlO$_x$ tunnel barrier approximately $1$-$2\,\mathrm{nm}$ thick, and a $5\,\mathrm{nm}$ top Al layer. These layers are covered by a $50\,\mathrm{nm}$ Nb counterelectrode (M2), and a $200\,\mathrm{nm}$ Nb wiring layer (M3) forms the upper-level interconnects. All thicknesses in this section are nominal deposition values based on previous calibrations. The scanning electron micrograph in the lower panel shows a tilted image of the resulting multilayer Josephson junction.

%\FloatBarrier
\section{Proximity Effect Modelling}
In the Nb/Al/AlO$_x$/Al/Nb stack, the thin Al interlayers are coupled to the outer Nb layers through the superconducting proximity effect~\cite{Zhao_1993_NbAl_proximity,Anferov_2024_Nb_Trilayer}, resulting in a superconducting gap that is much larger than that of bare Al while retaining the physical and chemical properties of the AlO$_x$ tunnel barrier. The superconducting gap can be modelled using the Usadel equations~\cite{Usadel_1970_generalized}, which are a set of diffusive-limit equations based on the Bardeen-Cooper-Schrieffer theory of superconductivity. 

Following Refs.~\cite{Brammertz_2001_proximity_model,Golubov_2004_current_phase,Songyuan_2018}, the equilibrium superconducting properties of each diffusive layer $S$ are parameterised using the spectral angle $\theta$. The one-dimensional Usadel equation is given by
\begin{equation} 
    \pi k_{\mathrm B} T_{\mathrm{c},S} \xi_S^2 \frac{\partial^2 \theta(x,\omega_n)}{\partial x^2} = \hbar\omega_n \sin\theta(x,\omega_n) - \Delta(x)\cos\theta(x,\omega_n), \label{eq:usadel_matsubara} 
\end{equation}
where $x$ is the coordinate normal to the film interfaces, $k_{\mathrm B}$ is the Boltzmann constant, $\hbar$ is the reduced Planck constant, $T_{\mathrm{c},S}$ is the intrinsic transition temperature of layer $S$, $\xi_S$ is the diffusive-limit coherence length of layer $S$, and $\Delta(x)$ is the spatially-dependent superconducting order parameter. The Matsubara frequencies are given by
\begin{equation}
    \omega_n
    =
    \frac{2\pi k_{\mathrm B}T}{\hbar}
    \left(n+\frac{1}{2}\right),
    \qquad
    n=0,1,2,\ldots\,,
    \label{eq:matsubara_frequencies}
\end{equation}
where $T$ is the bath temperature. The order parameter is then determined from the self-consistency equation
\begin{equation}
    \Delta(x)
    \ln\!\left(\frac{T_{\mathrm{c},S}}{T}\right)
    -
    2\pi k_{\mathrm B}T
    \sum_{\omega_n>0}
    \left[
        \frac{\Delta(x)}{\hbar\omega_n}
        -
        \sin\theta(x,\omega_n)
    \right]
    =0\,.
    \label{eq:self_consistency_matsubara}
\end{equation}

Following the self-consistent calculation of $\Delta(x)$, the real-energy spectral angle $\theta(x,E)$ is obtained through the analytic continuation $\hbar\omega_n\rightarrow-iE$. The resulting local single-spin quasiparticle density of states $N_S(x,E)$ is given by
\begin{equation}
    \frac{N_S(x,E)}{N_{0,S}}
    =
    \operatorname{Re}
    \left[
        \cos\theta(x,E)
    \right],
    \label{eq:local_dos}
\end{equation}
where $N_{0,S}$ is the single-spin normal-state density of states at the Fermi energy. 

At each free surface, the zero-current boundary condition is
\begin{equation}
    \frac{\partial\theta}{\partial x}=0.
    \label{eq:bc_open}
\end{equation}
For an interface between superconducting layers $S$ and $S'$, with the positive $x$ direction defined from $S$ towards $S'$, the Kupriyanov-Lukichev boundary conditions are given by~\cite{Kupriyanov_1988_boundary,Songyuan_2018}
\begin{equation}
    \begin{aligned}
        \sigma_S\frac{\partial\theta_S}{\partial x}
        &=
        \sigma_{S'}\frac{\partial\theta_{S'}}{\partial x},\\
        R_{\mathrm B}\sigma_S\frac{\partial\theta_S}{\partial x}
        &=
        \sin\!\left(\theta_{S'}-\theta_S\right).
    \end{aligned}
    \label{eq:bc_interface}
\end{equation}
The quantities $\theta_S$ and $\theta_{S'}$ denote the spectral angle on the two sides of the interface, $\sigma_S$ and $\sigma_{S'}$ are the corresponding normal-state conductivities, and $R_{\mathrm B}$ is the interface resistance-area product.

Conceptually, the Nb/Al/AlO$_x$/Al/Nb multilayer can be considered as two Nb/Al bilayers coupled through the AlO$_x$ tunnel barrier. To isolate the Nb-induced enhancement of superconductivity in Al, we analyse a single Nb/Al bilayer using the numerical Usadel-equation framework from Ref.~\cite{Songyuan_2018}. This captures the proximity-modified superconducting gap in the Al layer by the Nb layer, which is the principal effect. We model a single Nb/Al bilayer comprising a $120\,\mathrm{nm}$ Nb layer, a variable-thickness Al layer, and a clean Nb/Al interface described by Eq.~(\ref{eq:bc_interface}). The actual junction consists of two asymmetric Nb/Al electrodes with nominal Al thicknesses of $10\,\mathrm{nm}$ and $5\,\mathrm{nm}$, separated by an oxide tunnel barrier. The bilayer calculation therefore estimates the proximitisation of each electrode rather than modelling the complete junction stack. The material properties used in this simulation are summarised in Table~\ref{tab:usadelMaterialParameters}.

\begin{table}[htbp]
    \centering
    %\captionsetup{width=\textwidth}
    \caption{Material parameters used in the Nb/Al Usadel calculations.}%
    \label{tab:usadelMaterialParameters}
    \begin{tabular}{lcccc}
        \toprule
        Material & $T_{\mathrm{c}}$ (K) & $\sigma$ ($\mathrm{S\,m^{-1}}$) & $\xi_0$ (nm) & $\xi$ (nm) \\
        \midrule
        Al & $1.2^{\mathrm{\:a}}$ & $1.8\times10^{8\:\mathrm{a}}$ & $1600^{\mathrm{\:c}}$ & $147^{\mathrm{\:d}}$ \\
        Nb & $9.2^{\mathrm{\:b}}$ & $1.1\times10^{7\:\mathrm{b}}$ & $38^{\mathrm{\:c}}$ & $6.7^{\mathrm{\:e}}$ \\
        \bottomrule
    \end{tabular}

    \smallskip
    \footnotesize
    $^{\mathrm{a}}$ Al thin-film critical temperature and normal-state conductivity from Ref.~\cite{Songyuan_2018}.\par
    $^{\mathrm{b}}$ Nb thin-film critical temperature and normal-state conductivity from Ref.~\cite{Zhao_2023}.\par
    $^{\mathrm{c}}$ Bulk coherence length of Al and Nb from Ref.~\cite{Brammertz_2001_proximity_model}.\par
    $^{\mathrm{d}}$ Effective diffusive-limit coherence length of Al reported in Ref.~\cite{Brammertz_2001_proximity_model}, as estimated therein from the bulk Al coherence length and transmission-electron-microscopy measurements of the film grain size.\par
    $^{\mathrm{e}}$ Effective diffusive-limit coherence length of Nb reported in Ref.~\cite{Kushnir_2009_NbCu}, as estimated therein from upper critical field measurements.
\end{table}

\begin{figure}[htbp]
    \centering
    \begin{subfigure}[t]{0.45\textwidth}
        \centering
        \begin{tikzpicture}
            \node[anchor=south west,inner sep=0] (image) at (0,0) {\includegraphics[width=\textwidth]{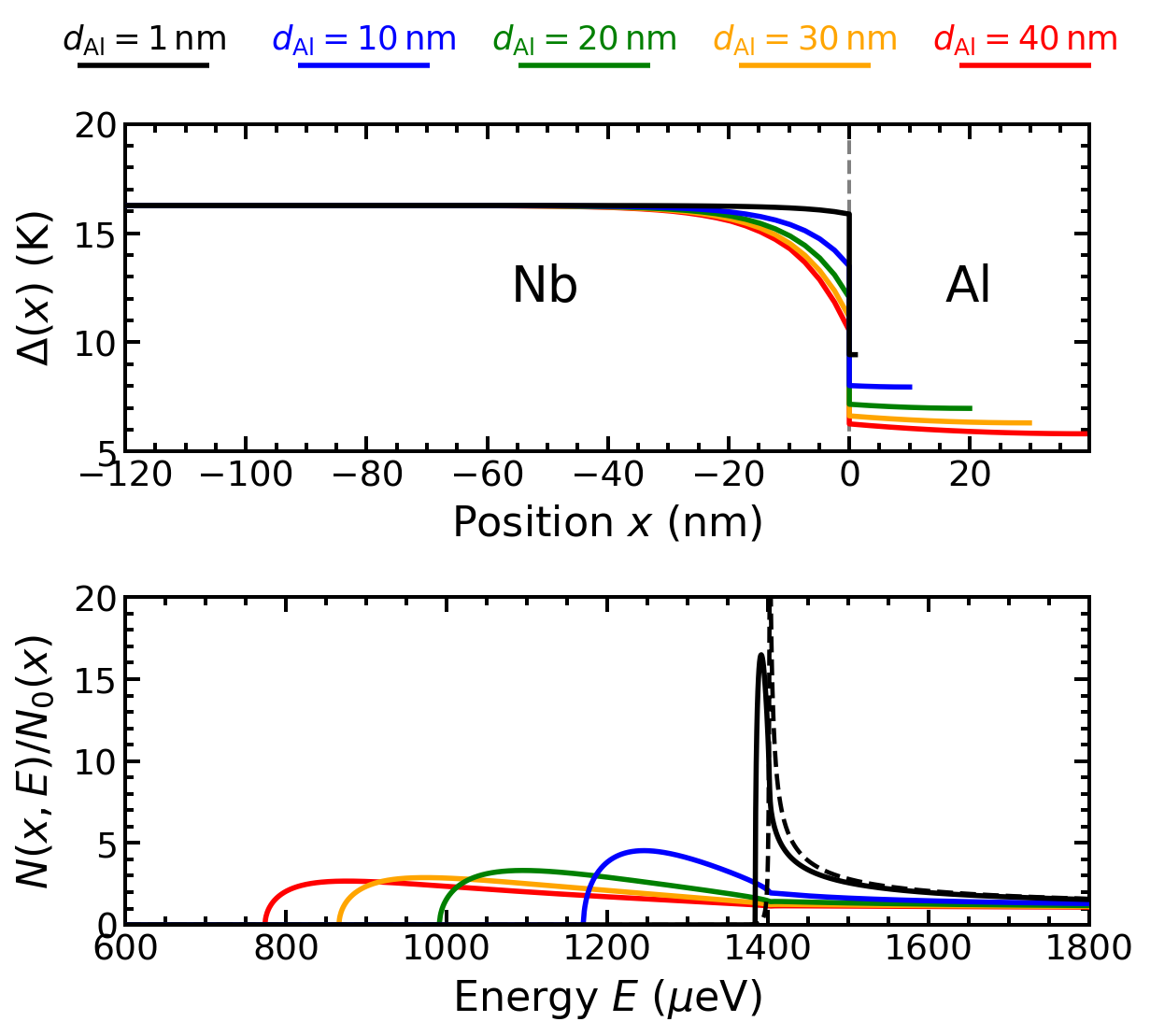}};
            \node[anchor=north west, text=black, font=\bfseries, xshift=-6pt, yshift=-6pt, overlay] at (image.north west) {a};
        \end{tikzpicture}
        \captionsetup{labelformat=empty, skip=0pt}%
        \caption{}%
        \label{fig:Usadel_simulation_profiles}
    \end{subfigure}\hfill
    \begin{subfigure}[t]{0.53\textwidth}
        \centering
        \begin{tikzpicture}
            \node[anchor=south west,inner sep=0] (image) at (0,0) {\includegraphics[width=\textwidth]{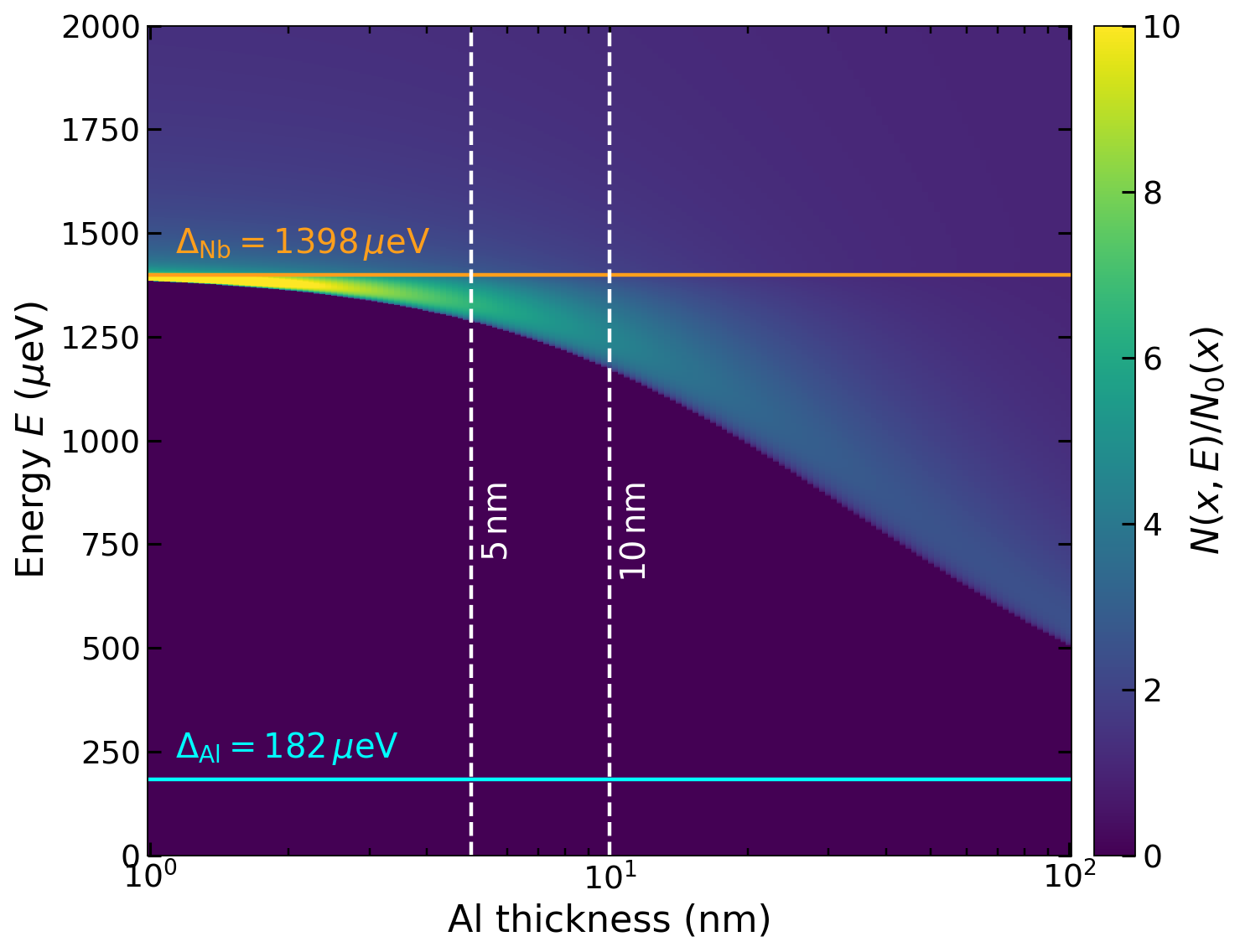}};
            \node[anchor=north west, text=black, font=\bfseries, xshift=-6pt, yshift=-6pt, overlay] at (image.north west) {b};
        \end{tikzpicture}
        \captionsetup{labelformat=empty, skip=0pt}%
        \caption{}%
        \label{fig:Usadel_simulation_map}
    \end{subfigure}
    \caption{\textbf{Numerical simulations of the self-consistent Usadel equations for a Nb/Al bilayer.} \textbf{a}~Spatially dependent superconducting order parameter, $\Delta(x)$, and normalised single-spin quasiparticle density of states, $N_S(x,E)/N_{0,S}$, for Al thicknesses of $1$, $10$, $20$, $30$, and $40\,\mathrm{nm}$. The density of states is evaluated at the outer surface of the Al layer, furthest from the Nb/Al interface, for all solid lines. The dashed black line shows the density of states at the outer Nb surface for the $1\,\mathrm{nm}$ Al bilayer. \textbf{b}~Density of states presented as a function of Al thickness and quasiparticle energy, illustrating the evolution of the proximity-modified energy gap. The orange and cyan lines indicate the bulk BCS gaps of Nb and Al, respectively. Vertical dashed lines indicate the nominal $5\,\mathrm{nm}$ and $10\,\mathrm{nm}$ Al layers used on opposite sides of the junction barrier.}%
    \label{fig:Usadel_simulation}
\end{figure}

Figure~\ref{fig:Usadel_simulation} shows numerical simulations of the self-consistent Usadel equations for a Nb/Al bilayer with a fixed Nb thickness of $120\,\mathrm{nm}$ and Al thicknesses ranging from $1\,\mathrm{nm}$ to $100\,\mathrm{nm}$ at $50\,\mathrm{mK}$. As shown in Fig.~\ref{fig:Usadel_simulation_profiles}, both the spatially dependent superconducting order parameter and the quasiparticle density of states are significantly modified by proximity coupling between Nb and Al, especially as the Al layer becomes thicker. The lower subfigure shows the quasiparticle density of states evaluated at the outer Al surface, plotted as solid lines. For the bilayer with a $1\,\mathrm{nm}$ Al layer, the black dashed line shows the density of states at the outer Nb surface for comparison. The solid and dashed black curves become non-zero at the same energy, indicating a common superconducting energy gap. This result is consistent with previous studies showing that proximity-coupled superconducting films up to several coherence lengths thick exhibit a common energy gap and critical temperature across their constituent layers~\cite{Brammertz_2001_proximity_model,Songyuan_2018,Zhao_2018}. The order parameter, in contrast, exhibits appreciable spatial dependence, especially near the Nb/Al interface. For small Al thicknesses on the order of a few nanometres, the density of states and the proximity-modified energy gap approach those of bulk Nb. For larger Al thicknesses, the density of states profile in the Al layer broadens and deviates substantially from the characteristic BCS profile.

The dependence of the superconducting energy gap on Al thickness is illustrated more clearly in Fig.~\ref{fig:Usadel_simulation_map}, which shows the collated densities of states at the outer Al surface for all Al thicknesses. The dark region at the bottom of the figure corresponds to energies at which the density of states vanishes, while the sharp onset of a finite density of states defines the superconducting energy gap, $\Delta_{\mathrm{g}}$. As the Al thickness increases, the outer Al surface becomes progressively less strongly coupled to Nb, and $\Delta_{\mathrm{g}}$ decreases towards the bulk Al value. At the Al thicknesses of approximately $5\,\mathrm{nm}$ and $10\,\mathrm{nm}$ used in this study, the calculated energy gaps remain approximately $92\%$ and $85\%$ of the bulk Nb gap, respectively, demonstrating strong proximitisation throughout the Al layers. In practical devices, Dynes broadening~\cite{Dynes_1978_lifetime_broadening}, which is not specific to proximity-coupled systems, may produce a weak tail of subgap states extending from the gap edge, potentially reducing the achievable quality factor and maximum operating temperature~\cite{Kubo_2022_subgap_states}. Depending on the sputtering conditions and film morphology, the conductivities and coherence lengths may also vary relative to the literature values used in these simulations.

\FloatBarrier
\section{Multilayer Qubit Characterisation}

\begin{figure}[ht]
    \centering
    \begin{subfigure}{0.49\textwidth}
        \centering
        \begin{tikzpicture}
            \node[anchor=south west,inner sep=0] (image) at (0,0) {\includegraphics[width=\textwidth]{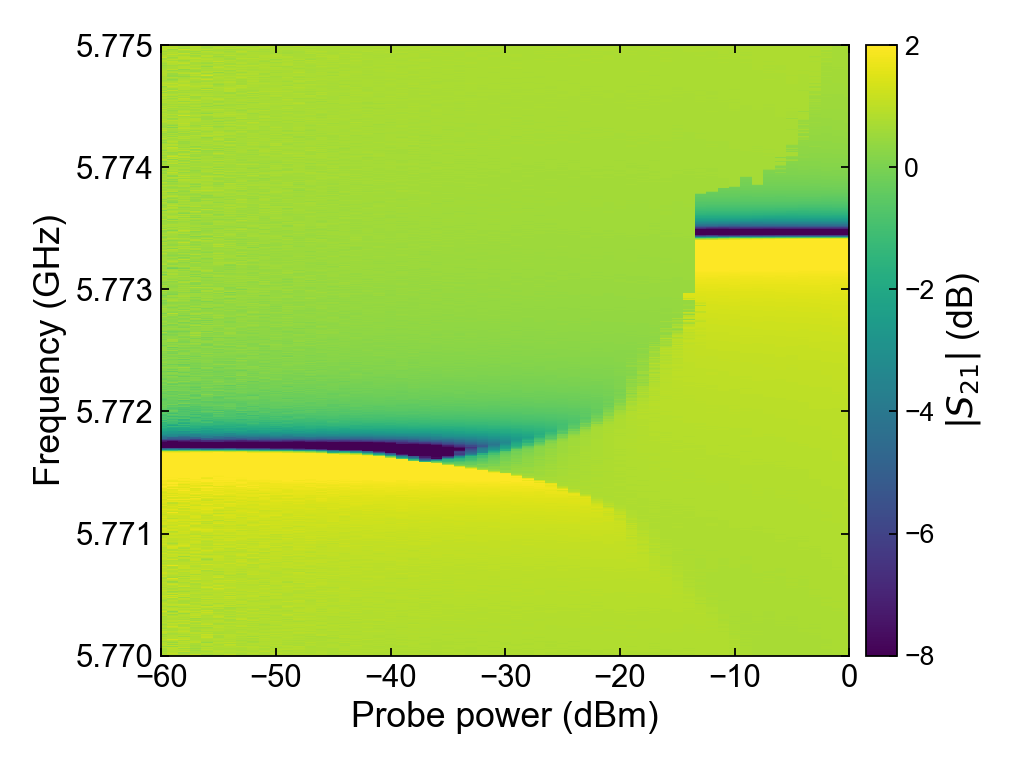}};
            \node[anchor=north west, text=black, font=\bfseries, xshift=-3pt, yshift=-6pt, overlay] at (image.north west) {a};
        \end{tikzpicture}
        \captionsetup{labelformat=empty, skip=0pt}%
        \caption{}%
        \label{fig:1_punchout}
    \end{subfigure}\hfill
    \begin{subfigure}{0.49\textwidth}
        \centering
        \begin{tikzpicture}
            \node[anchor=south west,inner sep=0] (image) at (0,0) {\includegraphics[width=\textwidth]{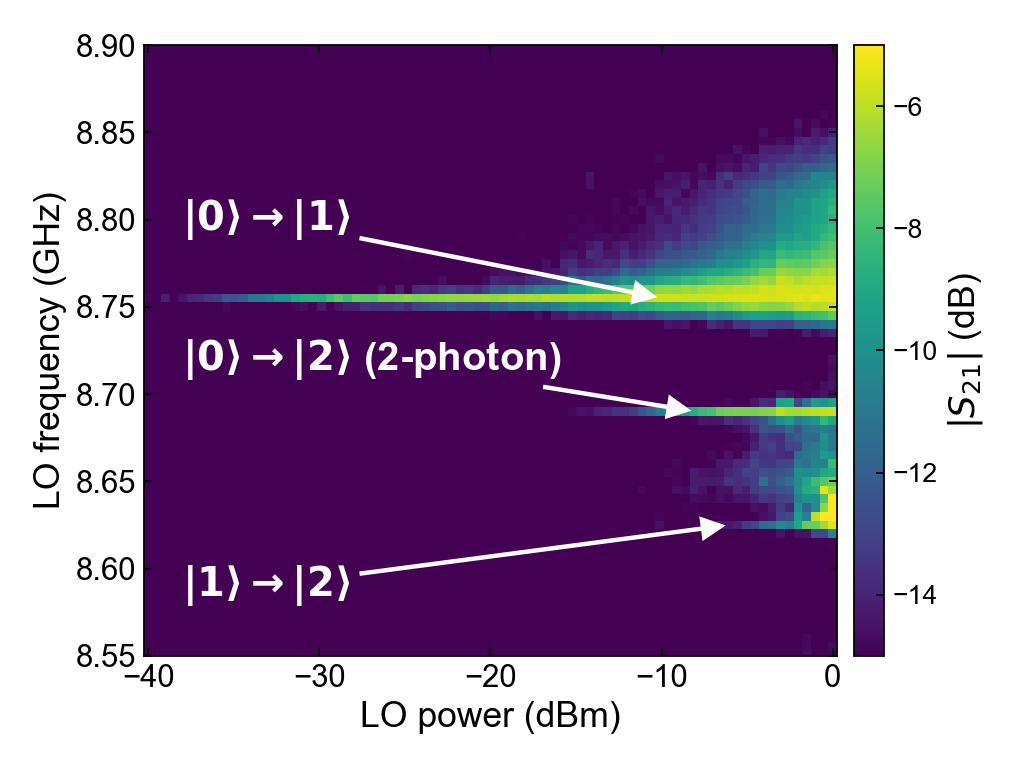}};
            \node[anchor=north west, text=black, font=\bfseries, xshift=-3pt, yshift=-6pt, overlay] at (image.north west) {b};
        \end{tikzpicture}
        \captionsetup{labelformat=empty, skip=0pt}%
        \caption{}%
        \label{fig:2_two_tone}
    \end{subfigure}
    
    \begin{subfigure}{0.49\textwidth}
        \centering
        \begin{tikzpicture}
            \node[anchor=south west,inner sep=0] (image) at (0,0) {\includegraphics[width=\textwidth]{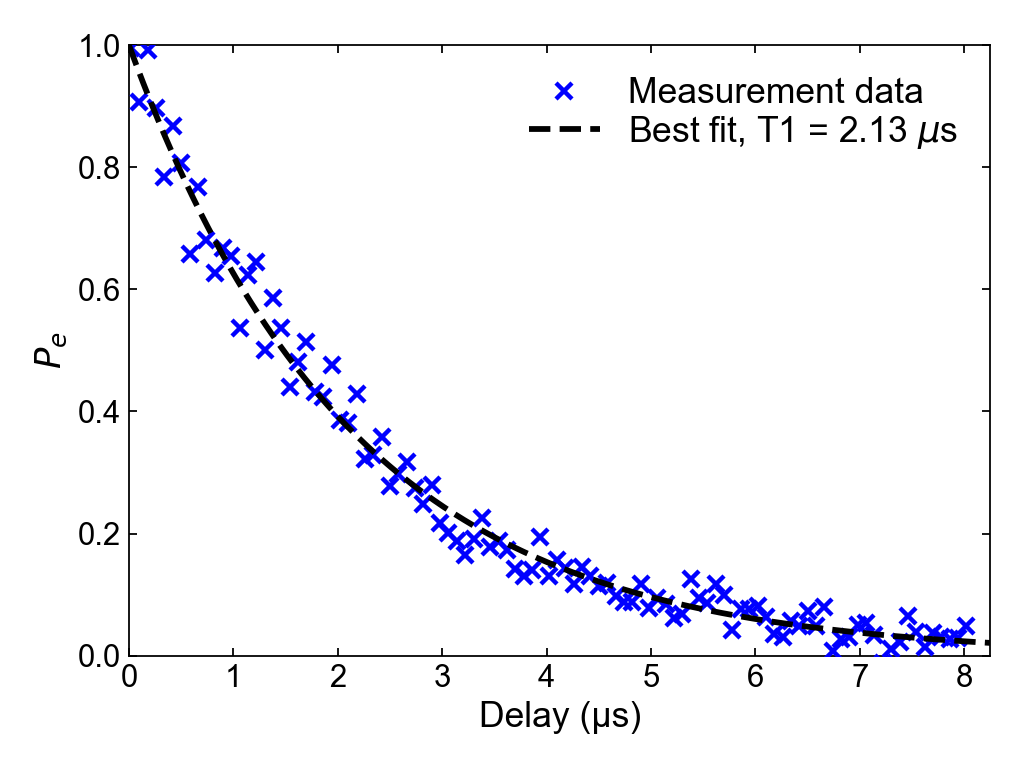}};
            \node[anchor=north west, text=black, font=\bfseries, xshift=-3pt, yshift=-6pt, overlay] at (image.north west) {c};
        \end{tikzpicture}
        \captionsetup{labelformat=empty, skip=0pt}%
        \caption{}%
        \label{fig:4_t1}
    \end{subfigure}\hfill
    \begin{subfigure}{0.49\textwidth}
        \centering
        \begin{tikzpicture}
            \node[anchor=south west,inner sep=0] (image) at (0,0) {\includegraphics[width=\textwidth]{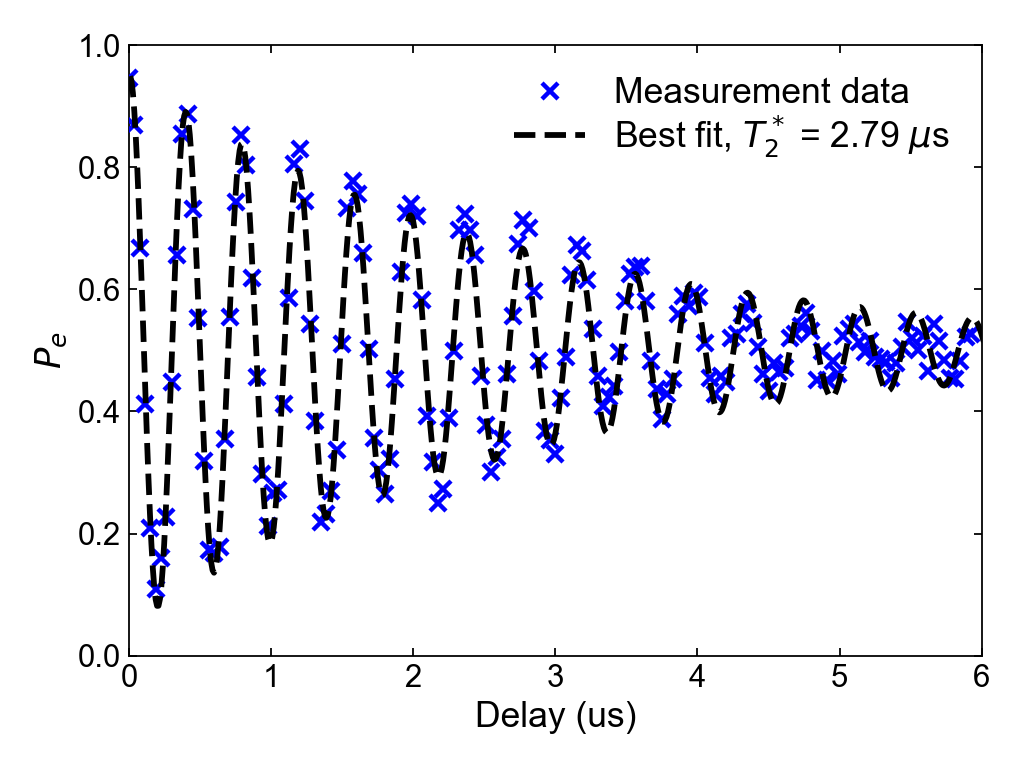}};
            \node[anchor=north west, text=black, font=\bfseries, xshift=-3pt, yshift=-6pt, overlay] at (image.north west) {d};
        \end{tikzpicture}
        \captionsetup{labelformat=empty, skip=0pt}%
        \caption{}%
        \label{fig:5_t2_star}
    \end{subfigure}
    
    \caption{\textbf{\boldmath Spectroscopy and time-domain characterisation at $50\,\mathrm{mK}$.} \textbf{a}~Punchout of the readout resonator. At a probe power of $-50\,\mathrm{dBm}$ the resonator is dispersively coupled to the qubit, yielding a resonance at $5.7717\,\mathrm{GHz}$; at probe powers above $-10\,\mathrm{dBm}$ the resonator decouples from the qubit, with the resonance shifting to $5.7735\,\mathrm{GHz}$. \textbf{b}~Two-tone spectroscopy of the qubit. The peak at $8.755\,\mathrm{GHz}$ corresponds to the $|0\rangle\rightarrow|1\rangle$ transition, that at $8.690\,\mathrm{GHz}$ to the two-photon $|0\rangle\rightarrow|2\rangle$ transition, and that at $8.625\,\mathrm{GHz}$ to the $|1\rangle\rightarrow|2\rangle$ transition. \textbf{c}~Energy relaxation $T_1$. \textbf{d}~Dephasing time $T_2^*$. The extracted $T_1$ and $T_2^*$ values lie in the ranges $1.5$-$2.5\,\mu\mathrm{s}$ and $2.0$-$3.0\,\mu\mathrm{s}$, respectively, depending on the measurement run.}%
    \label{fig:50mK_characterisation}
\end{figure}

We next characterised the Nb/Al/AlO$_x$/Al/Nb qubits using spectroscopic and time-domain measurements. All measurements in this study were performed in a dilution refrigerator at $50\,\mathrm{mK}$, unless otherwise stated. The cryogenic signal path and room-temperature measurement configurations are shown in Fig.~\ref{fig:cryostat_schematic} and Fig.~\ref{fig:measurement_schematic} (Supplementary Information Secs.~\ref{supp_cryostat} and~\ref{supp_measurements}). We did not incorporate parametric amplifiers in the readout chain, and relied on a low-noise high-electron-mobility transistor (HEMT) amplifier at $3\,\mathrm{K}$ for signal amplification. An adequate signal-to-noise ratio was ensured through time averaging in spectroscopy measurements by controlling the IF bandwidth of the vector network analyser (VNA), and through repeated-shot averaging in the time-domain measurements. All powers in this work are referenced at the point of generation, i.e., the output port of the VNA, signal generator, or arbitrary waveform generator (AWG).

Three resonator-qubit systems coupled to a common feedline were characterised at $50\,\mathrm{mK}$. The measured parameters for all devices are summarised in Table~\ref{tab:allDevices}, including the readout-resonator frequencies $f_{\mathrm{r}}$, qubit transition frequencies $f_{\mathrm{q}}$, punchout shifts $\Delta f_{\mathrm{PO}}$, anharmonicities $\alpha$, and measured relaxation and dephasing times $T_1$ and $T_2^*$. Unless otherwise stated, the results presented in the following sections correspond to Qubit~3 with a dressed resonator frequency $f_{\mathrm{r}}$ of $5.7717\,\mathrm{GHz}$ and a qubit transition frequency $f_{\mathrm{q}}$ of $8.755\,\mathrm{GHz}$. 

\begin{table}[htbp]
    \centering
    \caption{Measured properties of the three Nb/Al/AlO$_{x}$/Al/Nb qubit devices at $50\,\mathrm{mK}$. $T_1$ and $T_2^*$ are given as observed ranges across five measurement runs. Note that Qubit~1 is a tuneable qubit characterised near its operating sweet spot. Its relatively low transition frequency is intentional and does not reflect fabrication variability.}%
    \label{tab:allDevices}
    \begin{tabular}{lcccccc}
        \toprule
        & $f_{\mathrm{r}}$ (GHz) & $f_{\mathrm{q}}$ (GHz) & $\Delta f_{\mathrm{PO}}$ (MHz) & $\alpha$ (MHz) & $T_1$ ($\mu\mathrm{s}$) & $T_2^*$ ($\mu\mathrm{s}$) \\
        \midrule
        Qubit~1 & 4.68 & 5.500 & 2.4 & -140 & 4.0--7.0 & 4.5--5.5 \\
        Qubit~2 & 5.14 & 8.850 & 1.2 & -130 & 1.0--2.5 & 1.0--2.5 \\
        Qubit~3 & 5.77 & 8.755 & 1.8 & -130 & 1.5--2.5 & 2.0--3.0 \\
        \bottomrule
    \end{tabular}
\end{table}

The spectroscopy measurement setup was configured according to the upper sub-figure of Fig.~\ref{fig:measurement_schematic}. The result from the resonator punchout measurement is shown in Fig.~\ref{fig:1_punchout}. At a probe power of $-50\,\mathrm{dBm}$, the resonator is dispersively coupled to the qubit and resonates at $5.7717\,\mathrm{GHz}$. At probe powers above $-10\,\mathrm{dBm}$, the qubit is saturated and effectively decouples from the resonator, shifting the resonance to $5.7735\,\mathrm{GHz}$. The resulting dispersive punchout shift is $1.8\,\mathrm{MHz}$. The qubit transition frequency was determined by two-tone spectroscopy (Fig.~\ref{fig:2_two_tone}), which resolves a primary peak at $8.755\,\mathrm{GHz}$ corresponding to the $|0\rangle\rightarrow|1\rangle$ transition, alongside the two-photon $|0\rangle\rightarrow|2\rangle$ transition at $8.690\,\mathrm{GHz}$ and the $|1\rangle\rightarrow|2\rangle$ transition at $8.625\,\mathrm{GHz}$. The resulting anharmonicity of $-130\,\mathrm{MHz}$ is consistent with the design expectations.

The time-domain measurement setup was configured according to the lower sub-figure of Fig.~\ref{fig:measurement_schematic}. All time-domain measurements were obtained from repeated-shot averaging of the qubit state, typically 4096 or 8192 shots per data point. The qubit was controlled and measured using microwave pulses generated by an arbitrary waveform generator and upconverted to the qubit frequency using an IQ mixer. The control tone was applied to the qubit through the feedline and the weakly coupled resonator. It followed the same signal path as the readout pulse in the cryostat.

The results from time-domain characterisation are shown in the bottom subfigures of Fig.~\ref{fig:50mK_characterisation}. The energy relaxation time $T_1$ was extracted from the exponential decay of the qubit population following a $\pi$ pulse, as shown in Fig.~\ref{fig:4_t1}, and the dephasing time $T_2^*$ from the decay of the qubit coherence signal as shown in Fig.~\ref{fig:5_t2_star}. The extracted $T_1$ and $T_2^*$ values lie in the ranges $1.5$-$2.5\,\mu\mathrm{s}$ and $2.0$-$3.0\,\mu\mathrm{s}$, respectively, depending on the measurement run. The devices investigated in this work were not specifically optimised for coherence, and subsequent fabrication iterations have achieved improved coherence times. However, this does not affect the central novelty of the present study, i.e., the experimental demonstration and quantitative characterisation of a temperature-dependent dispersive response in Nb-based transmon circuits up to $2\,\mathrm{K}$.

\begin{figure}[H]
    \centering
    \begin{subfigure}[t]{0.49\textwidth}
        \centering
        \begin{tikzpicture}
            \node[anchor=south west,inner sep=0] (image) at (0,0) {\includegraphics[width=\textwidth]{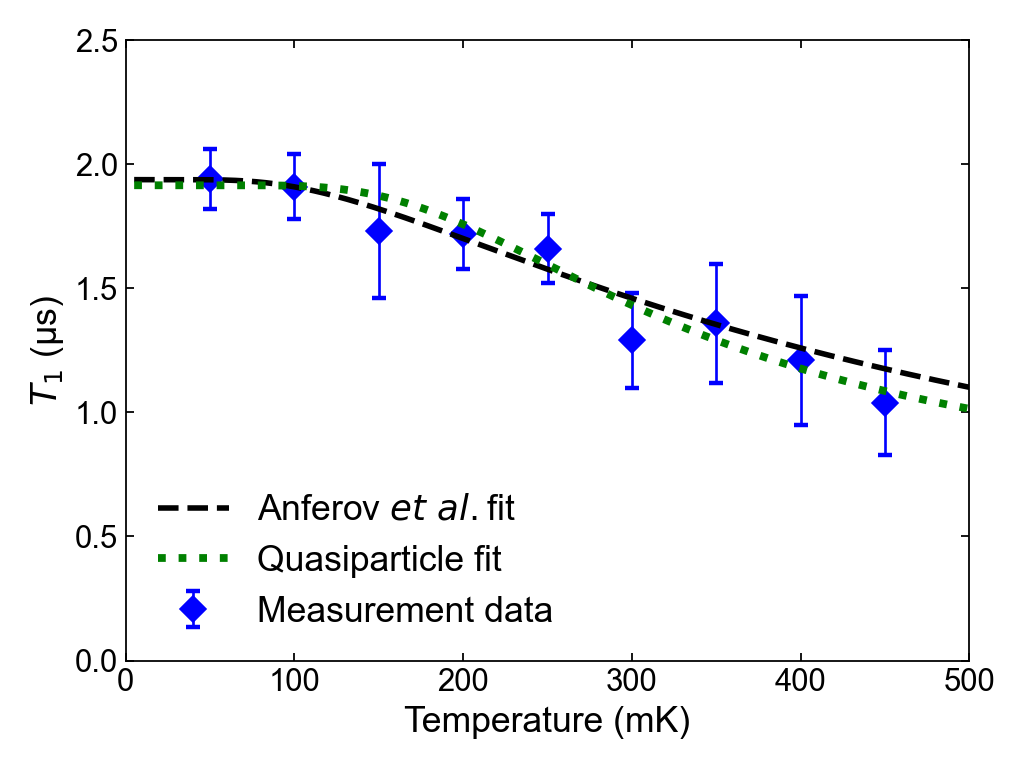}};
            \node[anchor=north west, text=black, font=\bfseries, xshift=-6pt, yshift=-6pt, overlay] at (image.north west) {a};
        \end{tikzpicture}
        \captionsetup{labelformat=empty, skip=0pt}%
        \caption{}%
        \label{fig:highT_T1}
    \end{subfigure}\hfill
    \begin{subfigure}[t]{0.49\textwidth}
        \centering
        \begin{tikzpicture}
            \node[anchor=south west,inner sep=0] (image) at (0,0) {\includegraphics[width=\textwidth]{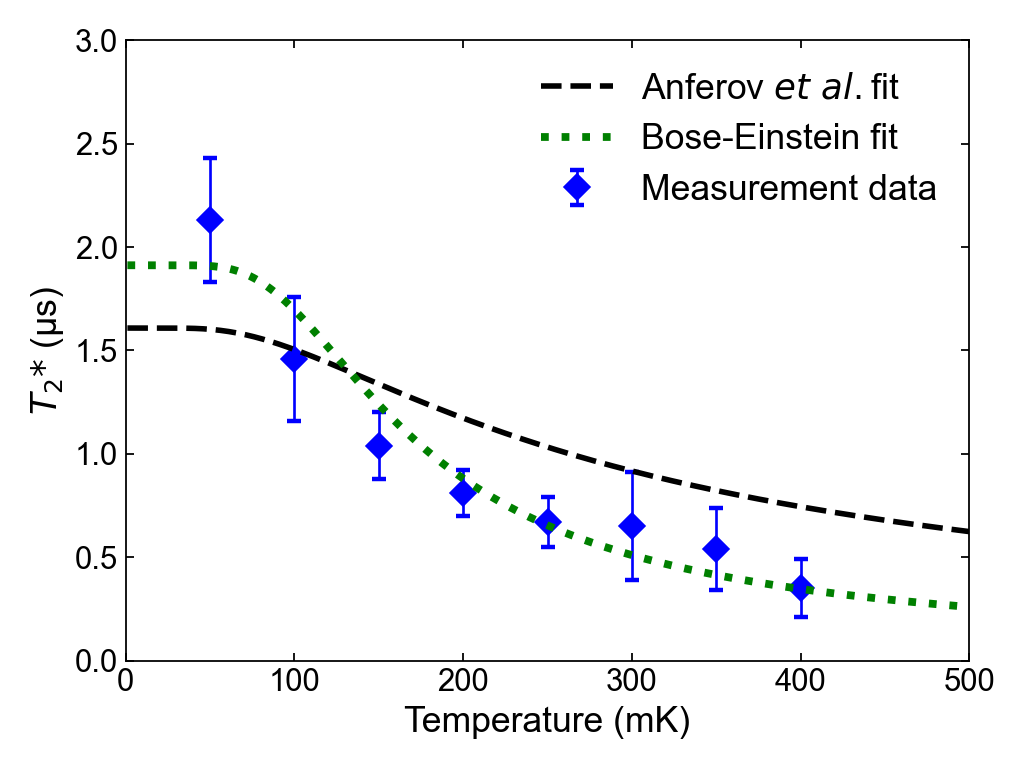}};
            \node[anchor=north west, text=black, font=\bfseries, xshift=-6pt, yshift=-6pt, overlay] at (image.north west) {b};
        \end{tikzpicture}
        \captionsetup{labelformat=empty, skip=0pt}%
        \caption{}%
        \label{fig:highT_T2}
    \end{subfigure}
    \caption{\textbf{\boldmath Temperature dependence of the qubit relaxation and Ramsey dephasing times.} All data points shown are averages of five measurements, with error bars representing the standard deviation.}%
    \label{fig:highT_T1_T2}
\end{figure}

In addition, time-domain characterisation was performed across the temperature range $50\,\mathrm{mK}$ to $400\,\mathrm{mK}$, as shown in Fig.~\ref{fig:highT_T1_T2}. All data points shown are averages of five measurements, with error bars representing the standard deviation. The qubit relaxation time $T_1$ shown in the left sub-figure can be well-fitted using the model in Appendix H of Ref.~\cite{Anferov_2024_20GHz_Qubits} derived from Refs.~\cite{Catelani_2011_quasiparticle_relaxation,Catelani_2014_parity_switching,Glazman_Catelani_2021_Bogoliubov_quasiparticles}, shown in the black dashed line. The underlying physics appears to be dominated by the quasiparticle population, governed by Eqs.~(1)~--~(2) of Ref.~\cite{Connolly_2024_quasiparticle_coexistence}, which likewise produce good agreement with the measured data as shown in the dotted green line.  The qubit dephasing time $T_2^*$ shown in the right sub-figure has been fitted against the Bose-Einstein population model according to Ref.~\cite{Sears_2012_photon_shot_noise}, shown in the dotted green line. The black dashed line shows a fit to the readout-dephasing model presented in Eq.~(1) and Appendix~H of Ref.~\cite{Anferov_2024_20GHz_Qubits}. The model describes dephasing caused by thermal photons in the readout resonator and is derived from Eq.~(44) of Ref.~\cite{Clerk_Utami_2007_photon_statistics}. The existing data and fitted models describe the observed temperature dependence within the resolution of the present measurement series.

\FloatBarrier
\section{Temperature Dependence of Dispersive Coupling}
We next investigated the temperature dependence of the dispersive coupling from $50\,\mathrm{mK}$ to $2\,\mathrm{K}$. While the temperature dependence of bare resonators has been extensively studied~\cite{Mattis_1958_anomalous_skin,Day_2003_broadband_detector,Pappas_2011_TLS_loss,Zmuidzinas_2012_microresonators}, that of the dispersively coupled resonance is less well explored, especially over the kelvin range investigated here. The temperature of our mixing-chamber plate was controlled via a stage heater, part of the PID system pre-integrated into the cryostat. After the PID loop was stabilised, the temperature was held constant for a further 15 minutes before measurements were taken to allow additional thermal settling. At each temperature, the resonator transmission was characterised as a function of probe frequency and input power, following the punchout procedure described above. This measurement allowed the dressed and decoupled resonance frequencies and Q-factors to be compared across the entire temperature range.

\begin{figure}[htbp]
    \centering
    \includegraphics[width=\textwidth]{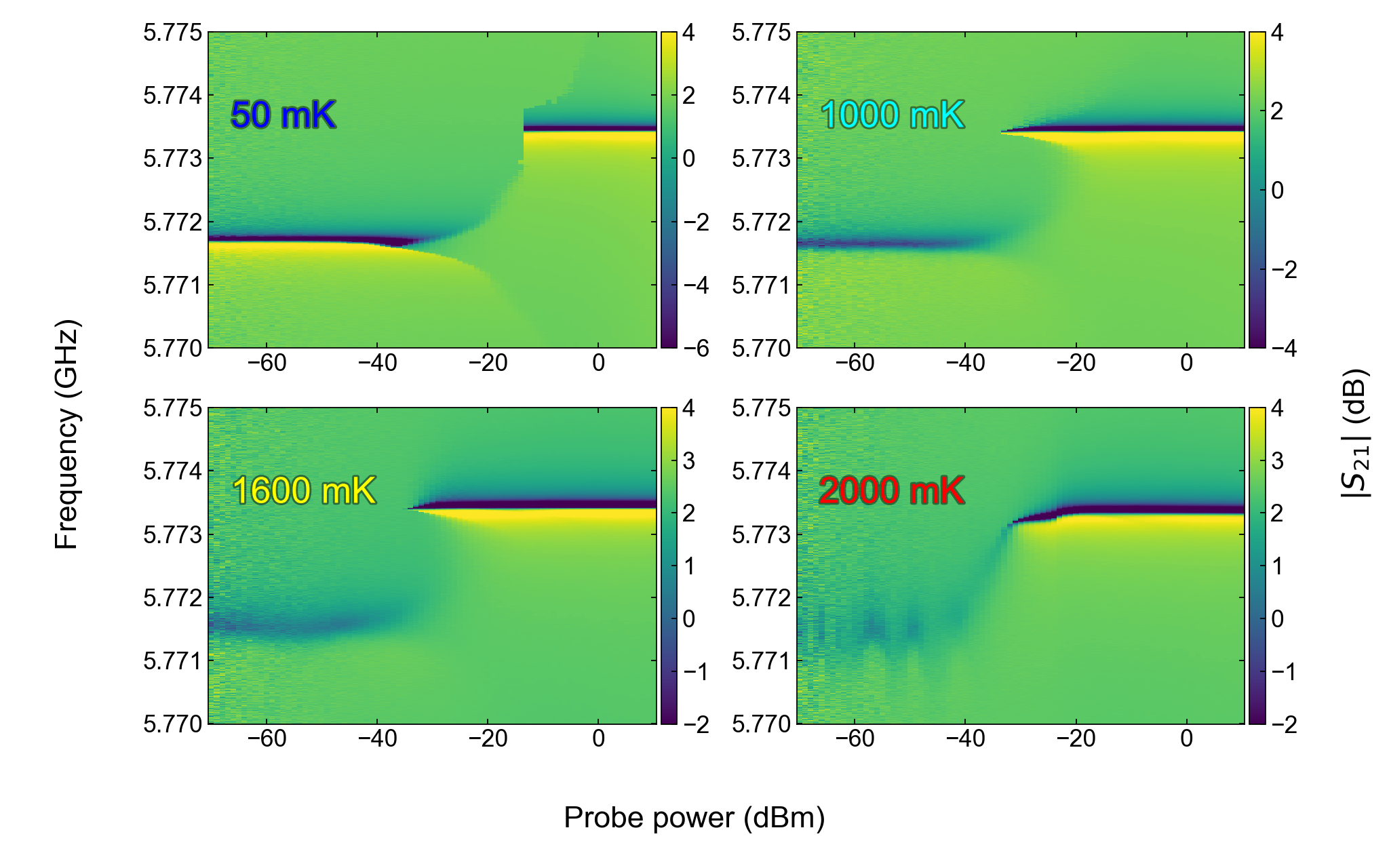}
    \caption{Punchout sweeps of the readout resonator at $50\,\mathrm{mK}$, $1\,\mathrm{K}$, $1.6\,\mathrm{K}$, and $2\,\mathrm{K}$. The probe powers are specified at the room-temperature VNA output rather than at the chip. The dispersive frequency shift between the low- and high-power resonances remains visible up to $2\,\mathrm{K}$, where the resonance becomes less well defined.}%
    \label{fig:highT_punchout}
\end{figure}

The punchout measurements are shown in Fig.~\ref{fig:highT_punchout}. A resolvable separation between the low-power, qubit-dressed resonance and the high-power, decoupled resonance is observed at $50\,\mathrm{mK}$, $1\,\mathrm{K}$, and $1.6\,\mathrm{K}$. At $2\,\mathrm{K}$, a power-dependent difference between the low-power and high-power resonator response is still identifiable, although the low-power resonance is less well defined.

\begin{figure}[htbp]
    \centering
    \begin{subfigure}[t]{0.49\textwidth}
        \centering
        \begin{tikzpicture}
            \node[anchor=south west,inner sep=0] (image) at (0,0) {\includegraphics[width=\textwidth]{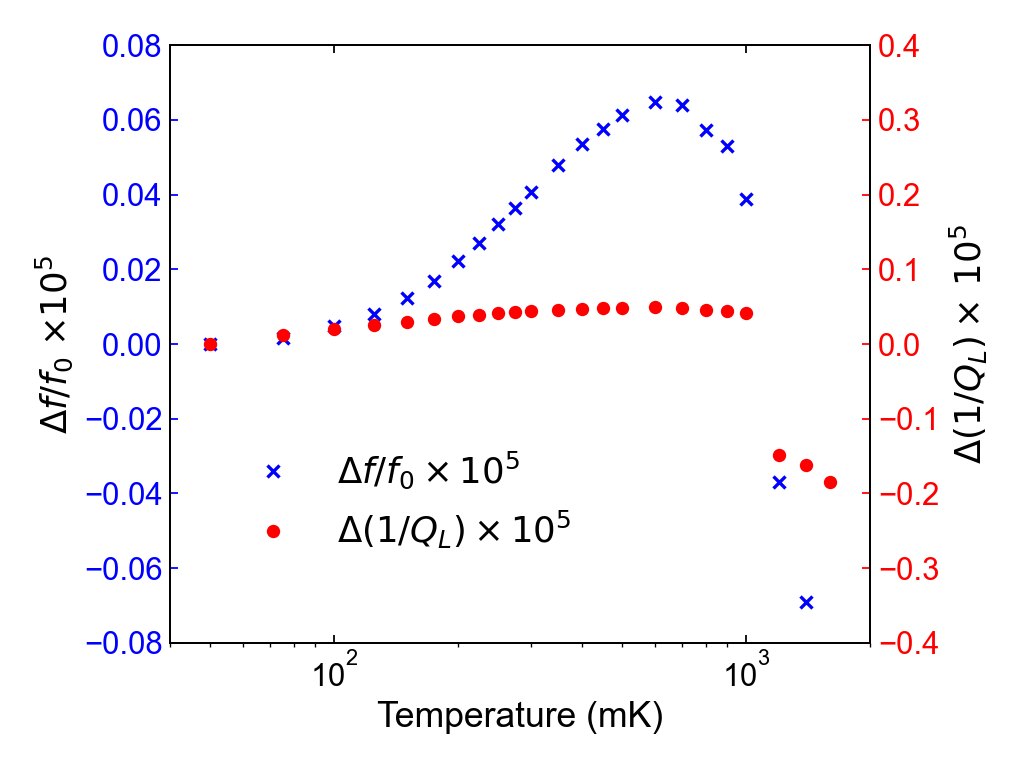}};
            \node[anchor=north west, text=black, font=\bfseries, xshift=-0pt, yshift=-6pt, overlay] at (image.north west) {a};
        \end{tikzpicture}
        \captionsetup{labelformat=empty, skip=0pt}%
        \caption{}%
        \label{fig:decoupled_resonator_shift}
    \end{subfigure}\hfill
    \begin{subfigure}[t]{0.49\textwidth}
        \centering
        \begin{tikzpicture}
            \node[anchor=south west,inner sep=0] (image) at (0,0) {\includegraphics[width=\textwidth]{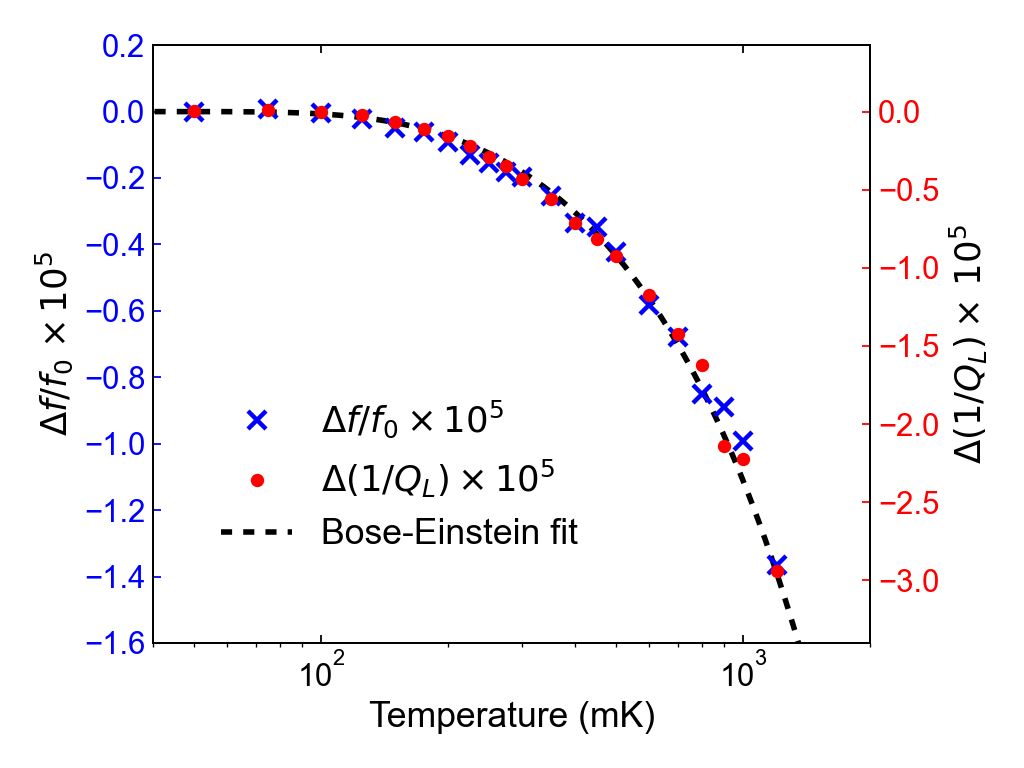}};
            \node[anchor=north west, text=black, font=\bfseries, xshift=-0pt, yshift=-6pt, overlay] at (image.north west) {b};
        \end{tikzpicture}
        \captionsetup{labelformat=empty, skip=0pt}%
        \caption{}%
        \label{fig:coupled_detuning}
    \end{subfigure}
    
    \begin{subfigure}[t]{0.49\textwidth}
        \centering
        \begin{tikzpicture}
            \node[anchor=south west,inner sep=0] (image) at (0,0) {\includegraphics[width=\textwidth]{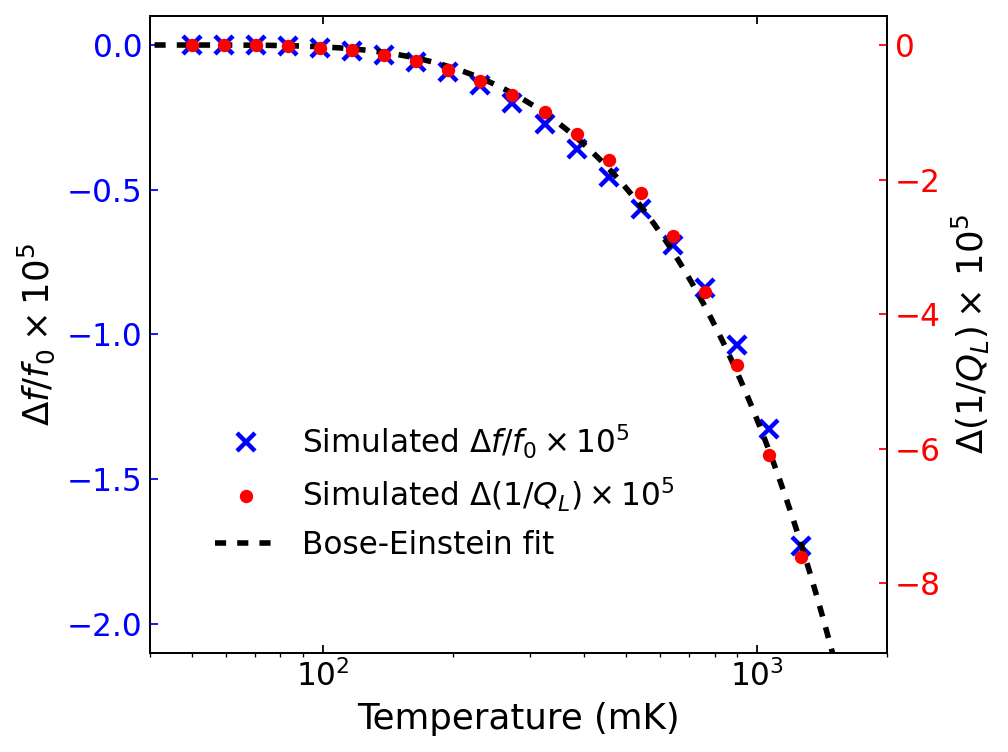}};
            \node[anchor=north west, text=black, font=\bfseries, xshift=-0pt, yshift=-6pt, overlay] at (image.north west) {c};
        \end{tikzpicture}
        \captionsetup{labelformat=empty, skip=0pt}%
        \caption{}%
        \label{fig:QuTiP_simulation}
    \end{subfigure}
    
    \caption{\textbf{\boldmath Characterisations of the decoupled and dressed readout resonances against temperature.} \textbf{a}~Fractional frequency detuning and change in inverse Q-factor of the decoupled readout resonance. \textbf{b}~Fractional frequency detuning and change in inverse Q-factor of the dressed-resonance. Blue crosses denote the dressed-resonance detuning, red circles show the change in inverse Q-factor, and the black dashed curve shows the best fit to the Bose-Einstein occupation. \textbf{c}~Temperature dependence of dressed-resonance detuning and inverse Q-factor for the Nb/Al/AlO$_x$/Al/Nb qubit simulated using QuTiP. The sign convention for the fractional frequency detuning is given by $\Delta f/f_0 = [f(T) - f_0]/f_0$, and the convention for the change in inverse loaded Q-factor is $\Delta(1/Q_L) = 1/Q_{L,0} - 1/Q_L(T)$, where $f_0$ and $Q_{L,0}$ are the $50\,\mathrm{mK}$ baseline values. Negative values of $\Delta f/f_0$ correspond to a downward shift in resonance frequency and negative changes in $\Delta(1/Q_L)$ correspond to increased loss.}%
    \label{fig:temperature_dependence}
\end{figure}

In order to quantitatively characterise the dispersive coupling, we extracted the resonance frequency and total Q-factor using the circle-fit procedure described in Ref.~\cite{Probst_2015_circle_fit} for both the dressed-resonance, measured at $-60\,\mathrm{dBm}$, and the decoupled resonance, measured at $0\,\mathrm{dBm}$. In Fig.~\ref{fig:temperature_dependence}, we plotted the temperature dependence of both quantities as differential changes relative to their baseline values at $50\,\mathrm{mK}$. Specifically, we define the fractional frequency detuning as $\Delta f/f_0 = [f(T) - f_0]/f_0$ and the change in inverse loaded Q-factor as $\Delta(1/Q_L) = 1/Q_{L,0} - 1/Q_L(T)$, where $f_0$ and $Q_{L,0}$ are the $50\,\mathrm{mK}$ baseline values. Under this sign convention, negative values of $\Delta f/f_0$ correspond to a downward shift in resonance frequency and negative changes in $\Delta(1/Q_L)$ correspond to increased loss and a decrease in $Q_L$ with increasing temperature. 

It is important to first separate the temperature-dependence of the resonator from the temperature-dependence of the qubit and/or dispersive coupling. This was done by comparing the relative magnitude of the temperature-dependence in the coupled and decoupled resonances. The changes in decoupled resonance frequency and inverse Q-factor are shown in Fig.~\ref{fig:decoupled_resonator_shift} and those of the dressed resonance are shown in Fig.~\ref{fig:coupled_detuning}. The temperature-dependent changes in the high-power decoupled reference resonance are approximately an order of magnitude smaller than those of the low-power dressed resonance. We therefore identify the temperature dependence of the resonator-qubit interaction as the dominant contribution to the observed response in the dressed resonance. The decoupled Q-factor exhibits a sharp decrease between $1\,\mathrm{K}$ and $1.2\,\mathrm{K}$. This behaviour is likely the result of Al bondwires transitioning to their normal state, thereby introducing additional loss into the system. Given the relatively small magnitude of this change, the effect is clearly resolved only in the decoupled Q-factor and is less readily discernible in the dispersively coupled resonance data.

The decoupled frequency detuning and Q-factor changes are not monotonic. The frequency detuning, in particular, initially increases with temperature before decreasing rapidly at higher temperatures. This behaviour is consistent with the frequency-shift model of Ref.~\cite{Crowley_2023_disentangling_losses}, which decomposes the total detuning into TLS-induced and quasiparticle-induced contributions, as described by Eqs.~(4)~--~(6) therein. The observed turnover is attributed to the interplay between these two mechanisms, with the quasiparticle-induced contribution becoming increasingly important at elevated temperatures. These TLS and quasiparticle contributions are likely still present in the pre-punchout dressed-resonance data. However, as noted earlier, their temperature-dependent variations are likely to be an order of magnitude smaller than those arising from the resonator-qubit interaction, which we discuss below.

The extracted dressed-resonance detuning and change in inverse Q-factor relative to the $50\,\mathrm{mK}$ baseline are shown in Fig.~\ref{fig:coupled_detuning}. The readout power is $-60\,\mathrm{dBm}$ at the output of the vector network analyser, corresponding to a resonator occupancy of approximately $4$--$8$ photons. The blue crosses show the dressed-resonance detuning, red circles show the change in inverse Q-factor, and the black dashed curve shows the best fit to the Bose-Einstein occupation, obtained using least-squares regression. The fitting function used to generate the curve is
\begin{equation}
    y(T)=\frac{A}{\exp\!\left(hf_{\mathrm{fit}}/k_{\mathrm{B}}T\right)-1},
    \label{eq:bose_einstein_fit}
\end{equation}
where the amplitude $A$ and the effective frequency $f_{\mathrm{fit}}$ are fitting parameters. The thermal occupation at the reference temperature of $50\,\mathrm{mK}$ is negligible within the experimental precision and is therefore omitted from the fitting expression. As shown in Fig.~\ref{fig:coupled_detuning}, the fit provides a good description of the frequency detuning data with an $R^2$ value of $0.991$, with a best-fit frequency $f_{\mathrm{fit}}=9.3\,\mathrm{GHz}$. Constraining $f_{\mathrm{fit}}$ to the resonator frequency $f_{\mathrm{r}} = 5.77\,\mathrm{GHz}$ still yields $R^2 = 0.985$. The close agreement between the measured points and the best-fit curve indicates that the temperature dependence of the extracted dressed detuning follows the thermal occupation of a bosonic mode. The fitting frequency $f_{\mathrm{fit}}$ characterises the effective temperature scale of the coupled response in frequency units. The fitted value of $f_{\mathrm{fit}}$ is close to the measured qubit transition frequency $f_{\mathrm{q}}=8.755\,\mathrm{GHz}$. Furthermore, as shown in the red scatter plot of Fig.~\ref{fig:coupled_detuning}, the change in the \textit{inverse Q-factor} of the dressed resonance likewise closely follows the Bose-Einstein occupation with an $R^2$ value of $0.976$ with respect to $f_{\mathrm{fit}}=9.3\,\mathrm{GHz}$ from the \textit{detuning} data. 

The correlated temperature dependences of the resonance detuning and inverse Q-factor are consistent with their interpretation as the dispersive and dissipative components of a common causal response connected by the Kramers-Kronig relations. This common response can be understood from the interplay between the occupation-dependent bosonic terms in the Jaynes-Cummings Hamiltonian and the thermal excitation and dissipation described by the Lindblad operators. The bosonic dependence of the Jaynes-Cummings Hamiltonian can be understood from the third-order expansion performed by Boissonneault \emph{et al.}~\cite{Boissonneault_2009_dispersive_regime}, in which Eq.~(3.11) gives the transformed Hamiltonian as
\begin{equation}
    H_s^D \simeq \hbar(\omega_r+\zeta)a^\dagger a
    +\hbar\left[\omega_q+2\chi\left(a^\dagger a+\frac{1}{2}\right)\right]\frac{\sigma_z}{2}
    +\hbar\zeta\left(a^\dagger a\right){}^2\sigma_z,
    \label{eq:dispersive_hamiltonian_third_order}
\end{equation}
where $\Delta=\omega_q-\omega_r$, $\lambda=g/\Delta$, $\chi=g^2(1-\lambda^2)/\Delta$, and $\zeta=-g^4/\Delta^3$. The final term is the leading nonlinear correction to the dispersive Hamiltonian. For a fixed qubit state with $\sigma_z=\pm1$, the energy required to add one resonator photon is obtained from the discrete derivative $[E_{\sigma_z}(n+1)-E_{\sigma_z}(n)]/\hbar$, giving
\begin{equation}
    \omega_{r,\sigma_z}(n)=\omega_r+\zeta+\sigma_z\left[\chi+\zeta(2n+1)\right].
    \label{eq:occupancy_dependent_resonator_frequency}
\end{equation}
As a result, to leading nonlinear order, the measured resonator transition frequency detuning is proportional to the occupancy $n$. For a resonator dominated by thermal occupancy, the temperature-dependent contribution to the frequency shift is therefore proportional to $\langle n\rangle\sim{n}_{\mathrm{B}}(\omega,T)={\left[\exp(\hbar\omega/k_{\mathrm{B}}T)-1\right]}^{-1}$. Temperature-independent contributions, such as those from the coherent readout drive, are absorbed into the reference resonance frequency at the lowest temperature, which is $50\,\mathrm{mK}$ in the above set of measurements. 

Dissipation and thermal excitation are described by the Gorini-Kossakowski-Sudarshan-Lindblad master equation~\cite{Lindblad_1976_dynamical_semigroup,Breuer_2002_open_systems},
\begin{equation}
    \dot{\rho}=-\frac{i}{\hbar}[H,\rho]+\sum_\mu\mathcal{D}[L_\mu]\rho,\quad \mathcal{D}[L]\rho=L\rho L^\dagger-\frac{1}{2}\{L^\dagger L,\rho\},
    \label{eq:Lindblad_model}
\end{equation}
with local thermal collapse operators $L_{r,\downarrow}=\sqrt{\kappa(n_r+1)}a$ and $L_{r,\uparrow}=\sqrt{\kappa n_r}a^\dagger$ for the resonator~\cite{Vaaranta_2025_local_bath} and analogous adjacent-level operators $L_{j,\downarrow}=\sqrt{\gamma_1(n_j+1)}\ell_j$ and $L_{j,\uparrow}=\sqrt{\gamma_1 n_j}\ell_j^\dagger$ with $\ell_j=\sqrt{j+1}|j\rangle\langle j+1|$ for the transmon~\cite{Cattaneo_2019_local_global, Breuer_2002_open_systems}, together with a pure-dephasing operator $L_{\phi}=\sqrt{2\gamma_\phi}\,\hat{n}_q$, where $j$ labels the transmon level, and $\kappa$, $\gamma_1$, and $\gamma_\phi$ are the temperature-independent bare resonator and transmon energy-decay rates and the pure-dephasing rate. The thermal occupation factors are given by $n_{r} = {\left[\exp\!\left(h f_r/k_{\mathrm{B}}T\right) - 1\right]}^{-1}$ and $n_{j} = {\left[\exp\!\left(h f_{j,j+1}/k_{\mathrm{B}}T\right) - 1\right]}^{-1}$ for the resonator and the transmon transitions, respectively. In addition to the nonlinear Jaynes-Cummings Hamiltonian, the temperature-dependence thus also enters through the Bose-Einstein occupations in the Lindblad operators, resulting in leading-order corrections proportional to $n_{r,j}(T)$. 

We numerically simulated the driven resonator-qubit system in QuTiP to confirm the physical origin of the observed temperature dependences. The QuTiP model evaluates the Gorini-Kossakowski-Sudarshan-Lindblad master equation with thermal collapse operators as described above, together with the multilevel Jaynes-Cummings Hamiltonian given by Eqs.~(34) and~(86) of Ref.~\cite{Blais_2021_cQED_review}, 
\begin{equation}
    \frac{H(t)}{\hbar}=\omega_r a^\dagger a+\omega_q b^\dagger b+\frac{\alpha}{2}b^{\dagger 2}b^2+g(a^\dagger b+a b^\dagger)+\epsilon_d\left(a^\dagger e^{-i\omega_d t}+a e^{i\omega_d t}\right),
    \label{eq:JC_model}
\end{equation}
where $a$ and $b$ are the resonator and transmon annihilation operators, $\alpha$ is the transmon anharmonicity, $g$ is the transverse coupling, and the last term is the coherent readout drive at $\omega_d$. The model was evaluated in the rotating frame to obtain a time-independent generator, following Sec.~VII.A of Ref.~\cite{Blais_2021_cQED_review}. The steady-state $\mathcal{L}\rho_{\mathrm{ss}}=0$ was solved using \texttt{qutip.steadystate} without time integration~\cite{Johansson_2012_QuTiP,Johansson_2013_QuTiP2}. For each temperature, $\langle a\rangle_{\mathrm{ss}}=\mathrm{Tr}(a\rho_{\mathrm{ss}})$ was converted to $S_{21}$ and analysed using the same circle-fit procedure described in Ref.~\cite{Probst_2015_circle_fit}. Further details of our QuTiP implementation, including the full set of numerical input parameters, are given in Supplementary Information Sec.~\ref{supp_Sim}.

The simulated detuning and inverse Q-factor are shown in Fig.~\ref{fig:QuTiP_simulation}. As seen in the figure, the simulation reproduces the Bose-Einstein temperature dependence, with a best-fit frequency of $f_{\mathrm{fit}}=9.97\,\mathrm{GHz}$ and an $R^2$ value of $0.996$ against the simulated detuning data. The simulation also replicates the close correspondence between the detuning and the change in inverse Q-factor. The model therefore reproduces the dominant temperature-dependent response of the coupled resonator-qubit system. Together with the substantially smaller temperature-dependent changes of the high-power reference resonator (Fig.~\ref{fig:decoupled_resonator_shift}), this supports the conclusion that the coupled-system dynamics dominate the measured behaviour.

It is important not to associate the Hamiltonian contribution exclusively with the frequency shift or the Lindblad contribution exclusively with dissipation. The response arising from either contribution can contain both dispersive and dissipative components. This distinction can be examined in QuTiP through two complementary modifications: (1) using the full Jaynes-Cummings Hamiltonian while fixing the thermal occupation factors at constant values in the Lindblad operators, or (2) replacing the Hamiltonian with its second-order dispersive expansion without the third-order terms in Eq.~\ref{eq:dispersive_hamiltonian_third_order}, while retaining the temperature-dependent occupation factors in the Lindblad operators. In both cases, the model still produces changes in both resonance detuning and inverse Q-factor with temperature.

\FloatBarrier
\section{Comparison with Al Qubits}

\begin{figure}[htbp]
    \centering
    \includegraphics[width=\textwidth]{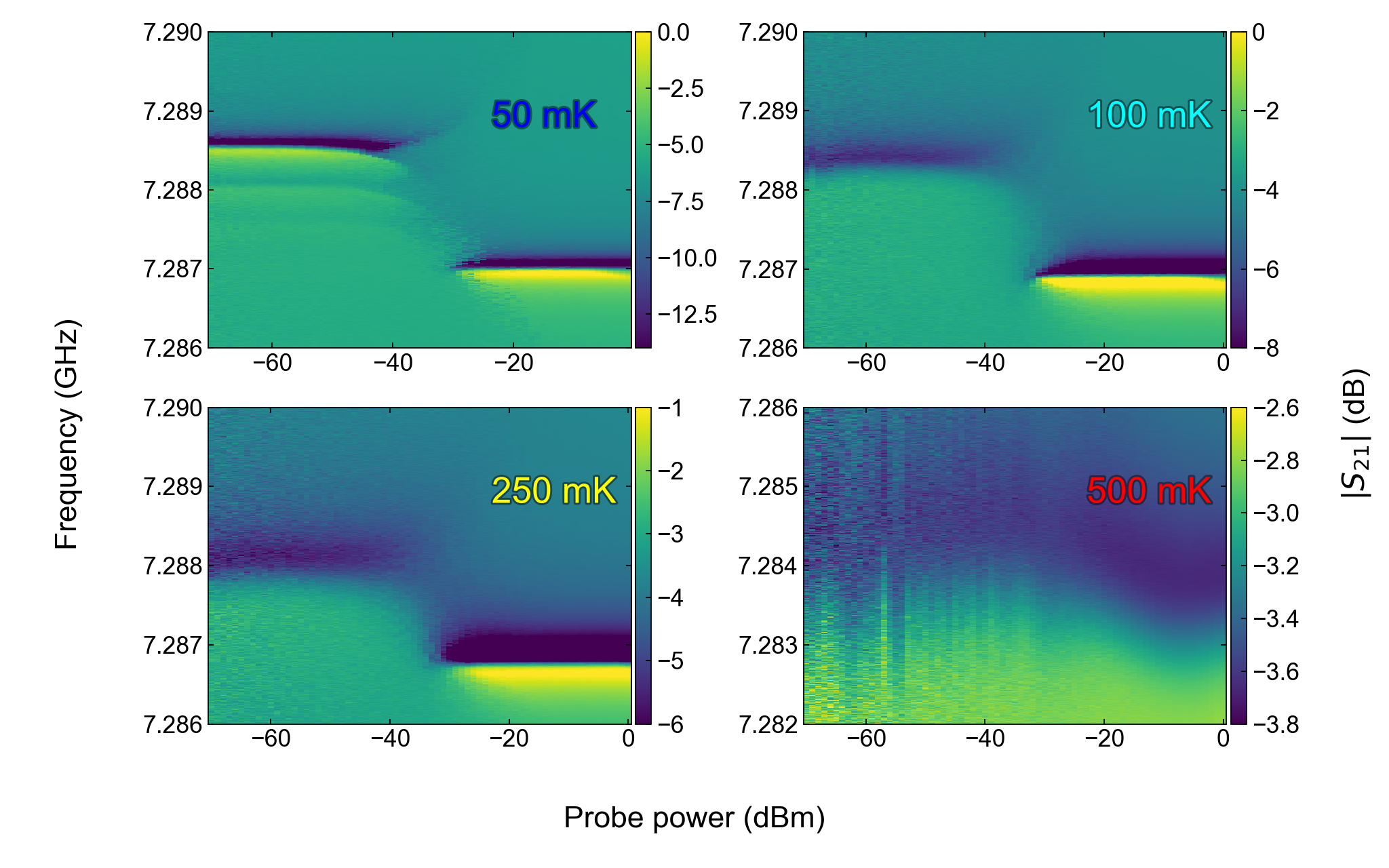}
    \caption{Temperature-dependent punchout sweeps of the Al readout resonator at $50\,\mathrm{mK}$, $100\,\mathrm{mK}$, $250\,\mathrm{mK}$, and $500\,\mathrm{mK}$. The dispersive separation between the low- and high-power resonances becomes progressively less stable with increasing temperature. At $500\,\mathrm{mK}$, the resonance shifts substantially and the dispersive response begins to break down.}%
    \label{fig:al_temperature_punchout}
\end{figure}

In order to establish a baseline reference for comparison with the Nb/Al/AlO$_x$/Al/Nb multilayer device, we also characterised an Al/AlO$_x$/Al transmon qubit coupled to an Al readout resonator at $50\,\mathrm{mK}$ using the same cryogenic and measurement setups. The Al qubit chip was fabricated by the National Physical Laboratory. The spectroscopic and time-domain characterisation measurements performed at $50\,\mathrm{mK}$ are summarised in Supplementary Information Sec.~\ref{supp_Al}.

The temperature-dependent Al punchout measurements are shown in Fig.~\ref{fig:al_temperature_punchout}. A clear dispersive response is observed at $50\,\mathrm{mK}$ and $100\,\mathrm{mK}$, while the separation between the dressed and decoupled resonances becomes less stable at $250\,\mathrm{mK}$. At $500\,\mathrm{mK}$, the resonance shifts significantly and the dispersive coupling begins to break down. 
The Al reference device therefore exhibits substantial degradation of its power-dependent dispersive response by $500\,\mathrm{mK}$, whereas the Nb-based multilayer device retains an identifiable dispersive signature up to $2\,\mathrm{K}$. This behaviour is consistent with the larger superconducting energy gap of the Nb-based electrodes compared with Al.

\FloatBarrier
\section{Conclusions}
We fabricated Nb/Al/AlO$_x$/Al/Nb multilayer transmons and investigated their temperature-dependent microwave response. Usadel-equation simulations with literature-derived thin-film parameters predict strong proximitisation of the Al interlayers. For a $120\,\mathrm{nm}$ Nb layer coupled to a $10\,\mathrm{nm}$ Al layer, the calculated energy gap in the Al layer reaches approximately $85\%$ of the bulk Nb gap. At $50\,\mathrm{mK}$, a transmon with a transition frequency of $8.755\,\mathrm{GHz}$ exhibits $T_1 = 1.5$--$2.5\,\mu\mathrm{s}$ and $T_2^* = 2.0$--$3.0\,\mu\mathrm{s}$, together with an anharmonicity of approximately $-130\,\mathrm{MHz}$ and a dispersive punchout shift of $1.8\,\mathrm{MHz}$. Existing time-domain measurements trace the reduction of these coherence times between $50$ and $400\,\mathrm{mK}$. Power-dependent resonator spectroscopy reveals an identifiable dispersive signature up to $2\,\mathrm{K}$, with deterioration of the low-power resonance at the highest temperatures, whereas an Al reference device measured in the same cryostat degrades substantially by $500\,\mathrm{mK}$. The measured changes in resonance frequency and inverse loaded Q-factor follow approximately Bose-Einstein-like thermal dependences, consistent with the response of a driven multilevel circuit-QED model. 

Thermal occupation at the present $5.77\,\mathrm{GHz}$ resonator and $8.755\,\mathrm{GHz}$ qubit frequencies remains substantial at $2\,\mathrm{K}$. Realising practical \textit{higher-temperature} operation will require \textit{higher-frequency} circuits together with suitable control, readout, and packaging. In particular, the development of suitable infrared filters~\cite{Hemmati_1985_absorbing_materials,Lamagna_2022_quasi_optical_testbed,Danilin_2022_noise_photon_flux} and parametric amplifiers~\cite{Shu_2021_NbTiN_parametric_amplification,Zhao_2026_NbN_resonator_paramps} at these frequencies remains an active and important area of research. Overall, the results in this present study provide a quantitative foundation for the development of higher-temperature and higher-frequency quantum architectures.

\FloatBarrier
\backmatter

\bmhead{Acknowledgements}
This work was supported by UKRI (SCALE-SQT, UKRI/EP/B000660/1; Super-CT, UKRI/EP/C001680/1; QCi3, EP/Z53318X/1; EPIQC, EP/W032627/1) and Innovate UK (FABU, grant 50868). We acknowledge the James Watt Nanofabrication Centre at the University of Glasgow for providing the nanofabrication infrastructure and technical expertise essential to this research. We thank Tobias Lindström and the National Physical Laboratory (NPL) for providing the aluminium qubit device studied in this work.

%\bmhead{Supplementary information}
\clearpage
\section*{Supplementary Information}
\renewcommand{\thesection}{S\arabic{section}}
\renewcommand{\thefigure}{S\arabic{figure}}
\setcounter{section}{0}
\setcounter{figure}{0}
\section{Cryostat configuration}\label{supp_cryostat}
\begin{figure}[tbp]
    \centering
    \includegraphics[width=0.70\textwidth]{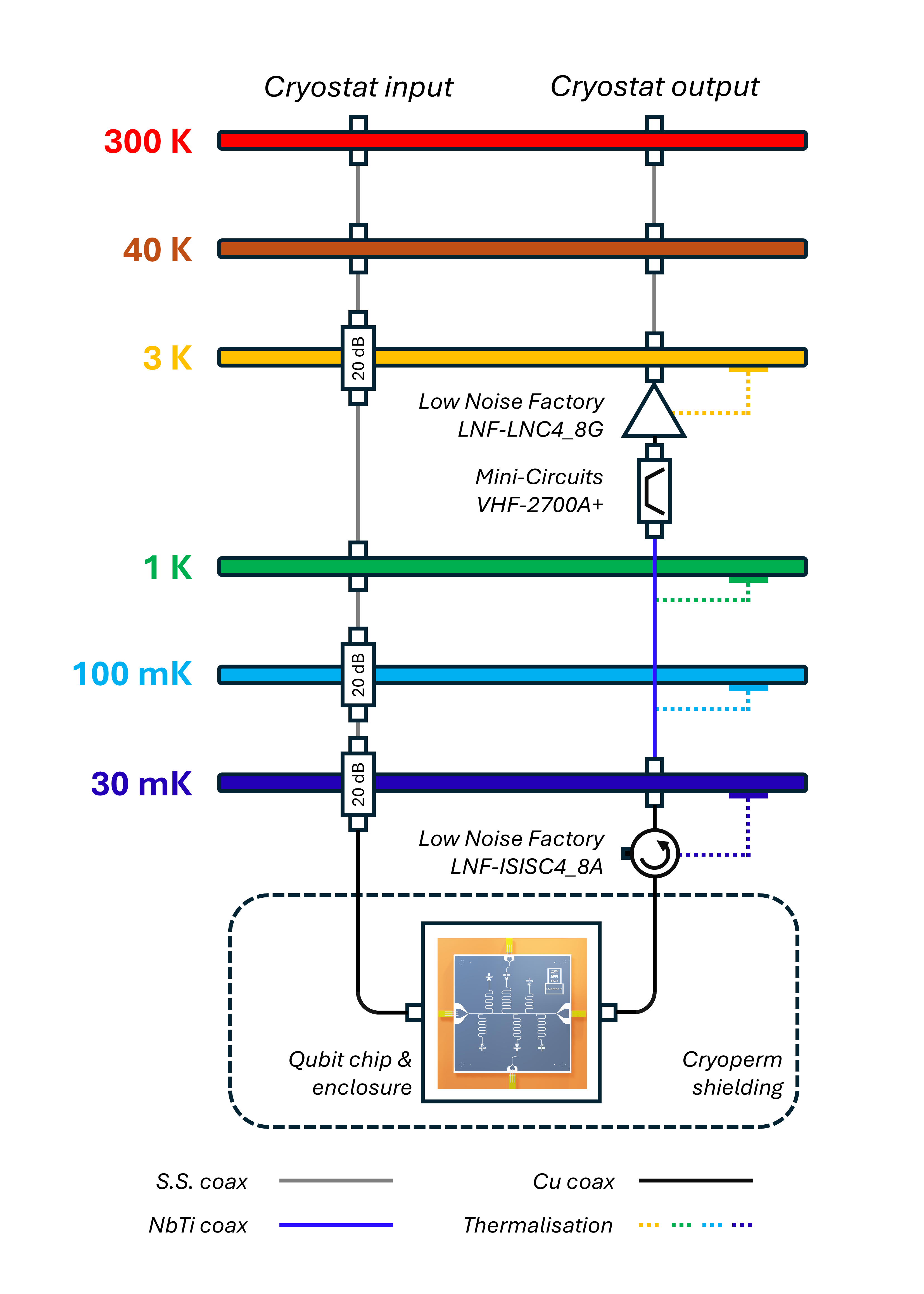}
    \caption{Schematic of the cryogenic wiring and thermalisation inside the dilution refrigerator.}\label{fig:cryostat_schematic}
\end{figure}
The measurements were performed in a dilution refrigerator configured according to Fig.~\ref{fig:cryostat_schematic}. The cryostat has temperature stages at $300\,\mathrm{K}$ (room-temperature stage), $40\,\mathrm{K}$, $3\,\mathrm{K}$, $1\,\mathrm{K}$, $100\,\mathrm{mK}$, and $30\,\mathrm{mK}$ (mixing-chamber plate). The input line is thermalised using $20\,\mathrm{dB}$ attenuators at the $3\,\mathrm{K}$, $100\,\mathrm{mK}$, and $30\,\mathrm{mK}$ stages to suppress room-temperature thermal noise and to ensure signal thermalisation~\cite{Krinner_2019_cryogenic_setups}. Stainless-steel coaxial cables are used for all stage transitions between the room-temperature stage and the mixing-chamber plate on the input side. Copper (Cu) coaxial cables are used to route signals between the sample and the mixing-chamber plate. 

The qubit chip and its enclosure are mounted at base temperature inside a high-permeability cryogenic magnetic shield. Although the lowest temperature is nominally $30\,\mathrm{mK}$, all data were acquired with the mixing chamber stabilised at $50\,\mathrm{mK}$ to avoid temperature fluctuations during the measurements.

On the output line, superconducting NbTi coaxial cables are used between the $3\,\mathrm{K}$, $1\,\mathrm{K}$, $100\,\mathrm{mK}$, and $30\,\mathrm{mK}$ stages. The output signal first passes through a Low Noise Factory LNF-ISISC4\_8A dual-junction isolator anchored at $30\,\mathrm{mK}$ before reaching a Low Noise Factory LNF-LNC4\_8G low-noise HEMT amplifier and a Mini-Circuits VHF-2700A+ high-pass filter at $3\,\mathrm{K}$, with thermalisation as indicated. 

\section{Measurement setup}\label{supp_measurements}
\begin{figure}[tbp]
    \centering
    \includegraphics[width=0.75\textwidth]{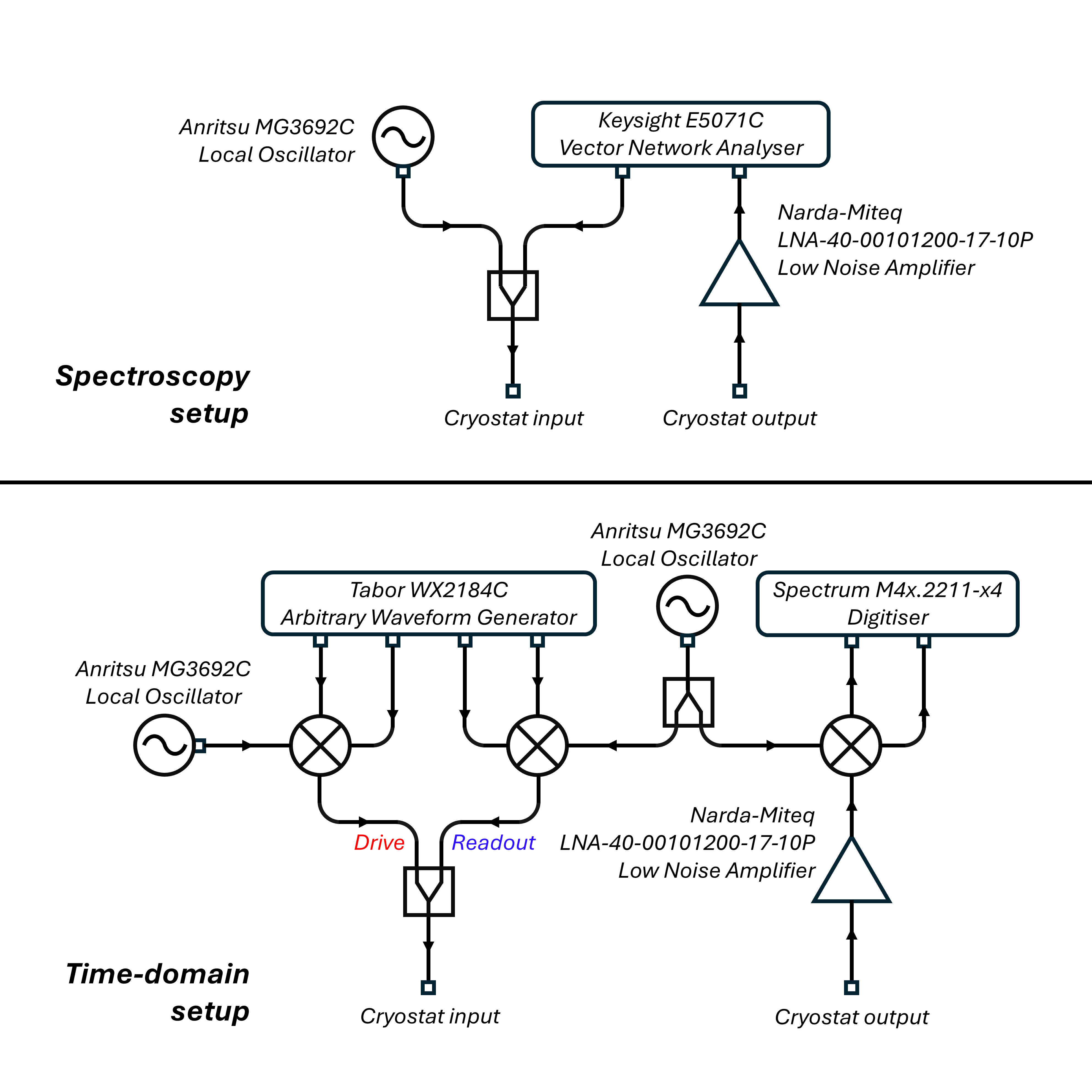}
    \caption{Schematic of the room-temperature measurement setups. Top: spectroscopy configuration. Bottom: time-domain control and heterodyne readout configuration.}\label{fig:measurement_schematic}
\end{figure}
Two room-temperature configurations were employed for spectroscopy and time-domain control and readout (Fig.~\ref{fig:measurement_schematic}). For spectroscopy measurements shown in the upper sub-figure, a Keysight E5071C vector network analyser provides the continuous-wave measurement tone, which is combined via a power combiner with a continuous-wave tone from an Anritsu MG3692C local oscillator before delivery to the cryostat input; the returning signal from the cryostat output is amplified by a Narda-Miteq LNA-40-00101200-17-10P low-noise amplifier before returning to the VNA for measurement of the complex transmission $S_{21}$.

For time-domain measurements shown in the lower sub-figure, baseband control and readout envelopes generated by a Tabor WX2184C arbitrary waveform generator are upconverted to microwave frequencies using IQ mixers driven by Anritsu MG3692C local oscillators, separately for the qubit drive and readout tones, and combined before delivery to the cryostat input; the output from the cryostat is amplified by a Narda-Miteq LNA-40-00101200-17-10P amplifier, downconverted by an IQ mixer driven by an Anritsu MG3692C local oscillator, and digitised using a Spectrum M4x.2211-x4 digitiser.

\section{Simulation parameters}\label{supp_Sim}
\begin{table}[htbp]
    \centering
    \caption{Device parameters used in the thermal master-equation simulation.}%
    \label{tab:modelParams}
    \setlength{\tabcolsep}{6pt}
    \begin{tabular}{cccccccc}
        \toprule
        $f_{\mathrm{r}}$ & $f_{\mathrm{q}}$ & $E_C/h$ & $g/2\pi$ & $\gamma_1/2\pi$ & $\gamma_\phi/2\pi$ & $Q_i$ & $Q_{L}$ \\
        (GHz) & (GHz) & (MHz) & (MHz) & (kHz) & (kHz) & ($10^{6}$) & ($10^{4}$) \\
        \midrule
        5.7735 & 8.755 & 130 & 73.3 & 80 & 25 & 1.45 & 6.40 \\
        \bottomrule
    \end{tabular}
\end{table}

The parameters in Table~\ref{tab:modelParams} are chosen to represent the properties of Qubit~3 given in Table~\ref{tab:allDevices}. The transmon and resonator Hilbert spaces were truncated to $N_{\mathrm{q}} = 8$ transmon levels and $N_{\mathrm{r}} = 12$ resonator Fock states, respectively. Convergence with respect to the Hilbert-space truncation can be assessed from the fraction of the thermal distribution retained. At $T = 1\,\mathrm{K}$, the truncations retain over $95\%$ of the resonator and transmon populations. The convergence was also verified by comparing the results obtained with $N_{\mathrm{q}} = 8$, $N_{\mathrm{r}} = 12$ against those with $N_{\mathrm{q}} = 9$, $N_{\mathrm{r}} = 13$, which retain approximately $97\%$ of the resonator and transmon populations. The values of $\Delta f/f_0$ and $\Delta(1/Q_{L})$ at $T = 1\,\mathrm{K}$ from the two numerical settings agree to within $5\%$ of each other, below the combined experimental uncertainties and negligible on the scale of the measured temperature dependence. The drive amplitude was chosen to be small compared to the thermal occupations, corresponding to $0.01$ photons. For each drive frequency, the cavity amplitude obtained from the QuTiP solver is converted to $S_{21}$. The resulting $S_{21}$ spectrum is then fitted using the same circle-fit procedure as the experimental data analysis, described in Ref.~\cite{Probst_2015_circle_fit}. An implementation of the circle-fit algorithm is available in the open-source Qkit repository~\cite{qkit_2024}.

\section{Pure-dephasing times}\label{supp_TPhi}
\begin{figure}[H]
    \centering
    \includegraphics[width=0.6\textwidth]{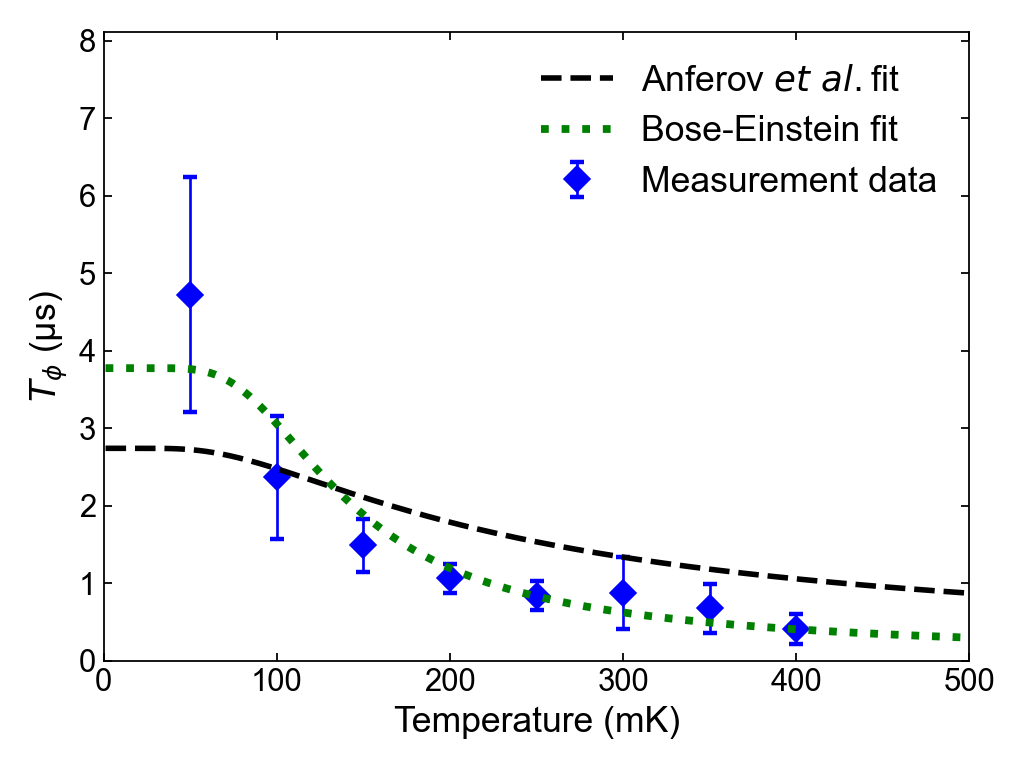}
    \caption{\textbf{\boldmath Temperature dependence of the qubit pure-dephasing times $T_\phi$, extracted using $1/T_2^*=1/(2T_1)+1/T_\phi$.}}%
    \label{fig:highT_Tphi}
\end{figure}

The effective pure-dephasing time $T_\phi$ is extracted from temperature-dependent $T_1$ and $T_2^*$ data using $1/T_2^*=1/(2T_1)+1/T_\phi$ from Ref.~\cite{Burnett_2019_decoherence_benchmarking}, and is fitted against the Bose-Einstein population model according to Eq.~(5) of Ref.~\cite{Yan_2016_flux_qubit_revisited} and against the same readout-dephasing model reported in Ref.~\cite{Anferov_2024_20GHz_Qubits}. 

\section{Al reference-qubit characterisation}\label{supp_Al}
\begin{figure}[H]
    \centering
    \begin{subfigure}[t]{0.49\textwidth}
        \centering
        \begin{tikzpicture}
            \node[anchor=south west,inner sep=0] (image) at (0,0) {\includegraphics[width=\textwidth]{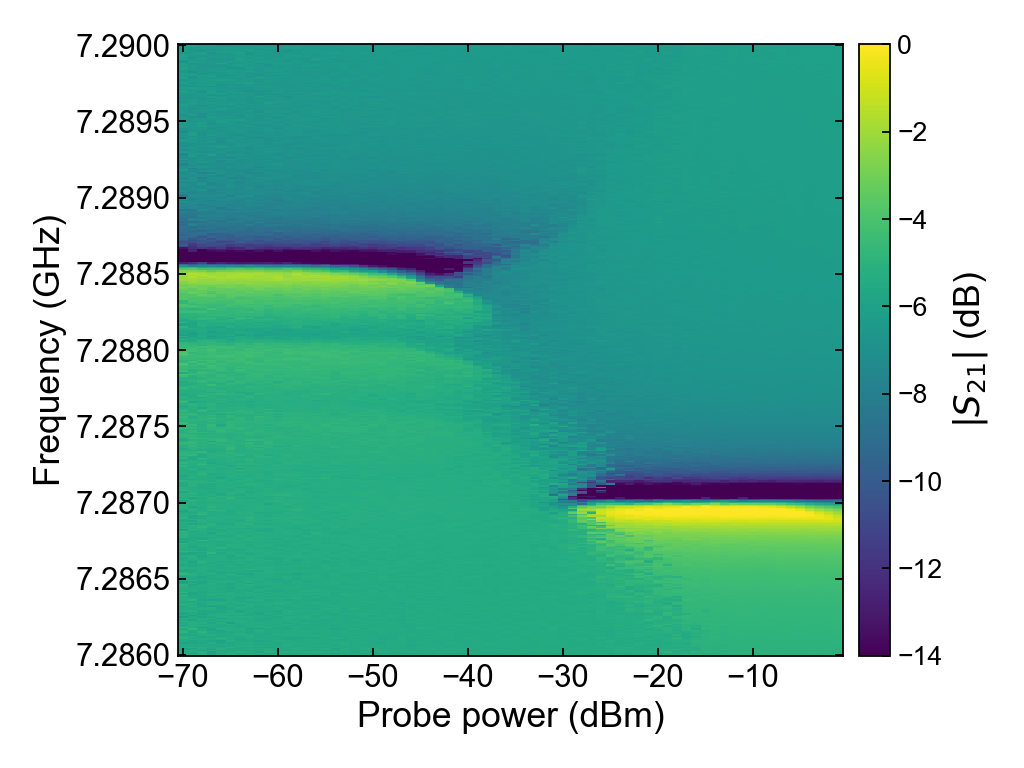}};
            \node[anchor=north west, text=black, font=\bfseries, xshift=-6pt, yshift=-6pt, overlay] at (image.north west) {a};
        \end{tikzpicture}
        \captionsetup{labelformat=empty, skip=0pt}%
        \caption{}%
        \label{fig:al_punchout}
    \end{subfigure}\hfill
    \begin{subfigure}[t]{0.49\textwidth}
        \centering
        \begin{tikzpicture}
            \node[anchor=south west,inner sep=0] (image) at (0,0) {\includegraphics[width=\textwidth]{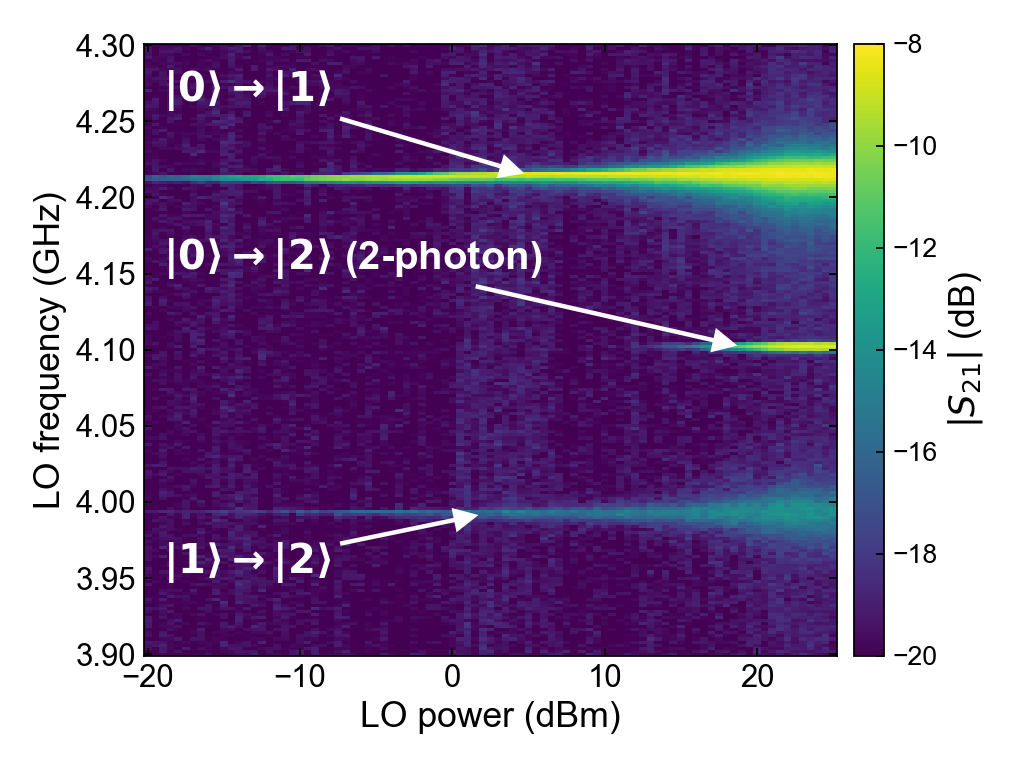}};
            \node[anchor=north west, text=black, font=\bfseries, xshift=-6pt, yshift=-6pt, overlay] at (image.north west) {b};
        \end{tikzpicture}
        \captionsetup{labelformat=empty, skip=0pt}%
        \caption{}%
        \label{fig:al_two_tone}
    \end{subfigure}
    \\
    \begin{subfigure}[t]{0.49\textwidth}
        \centering
        \begin{tikzpicture}
            \node[anchor=south west,inner sep=0] (image) at (0,0) {\includegraphics[width=\textwidth]{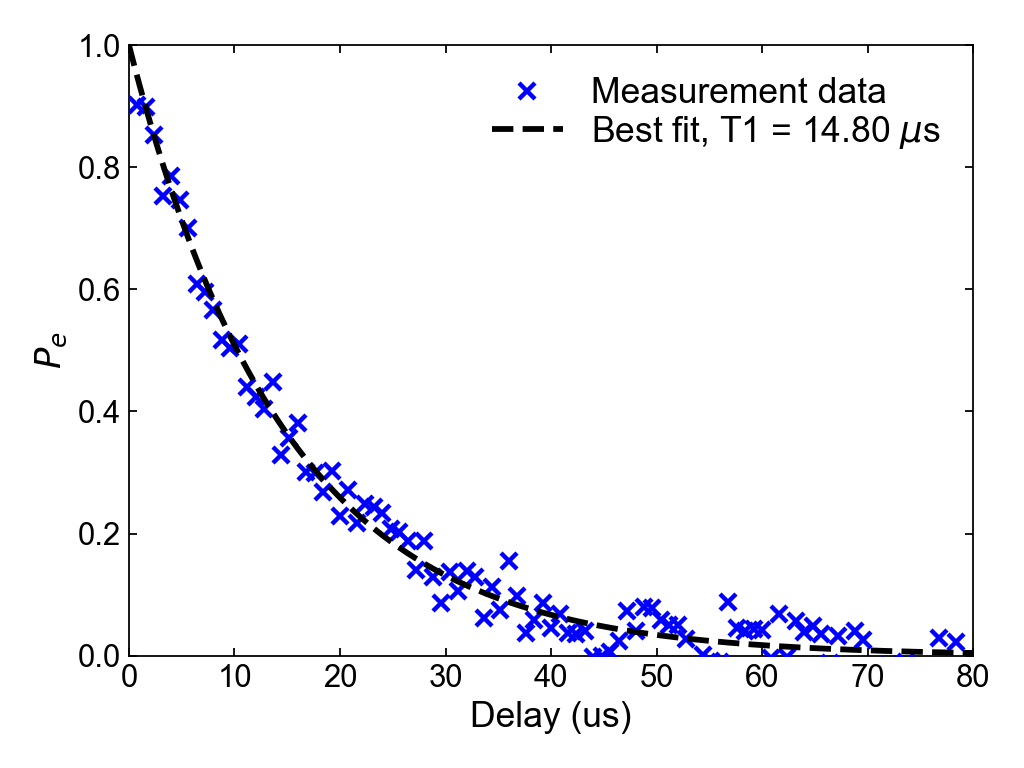}};
            \node[anchor=north west, text=black, font=\bfseries, xshift=-6pt, yshift=-6pt, overlay] at (image.north west) {c};
        \end{tikzpicture}
        \captionsetup{labelformat=empty, skip=0pt}%
        \caption{}%
        \label{fig:al_t1}
    \end{subfigure}\hfill
    \begin{subfigure}[t]{0.49\textwidth}
        \centering
        \begin{tikzpicture}
            \node[anchor=south west,inner sep=0] (image) at (0,0) {\includegraphics[width=\textwidth]{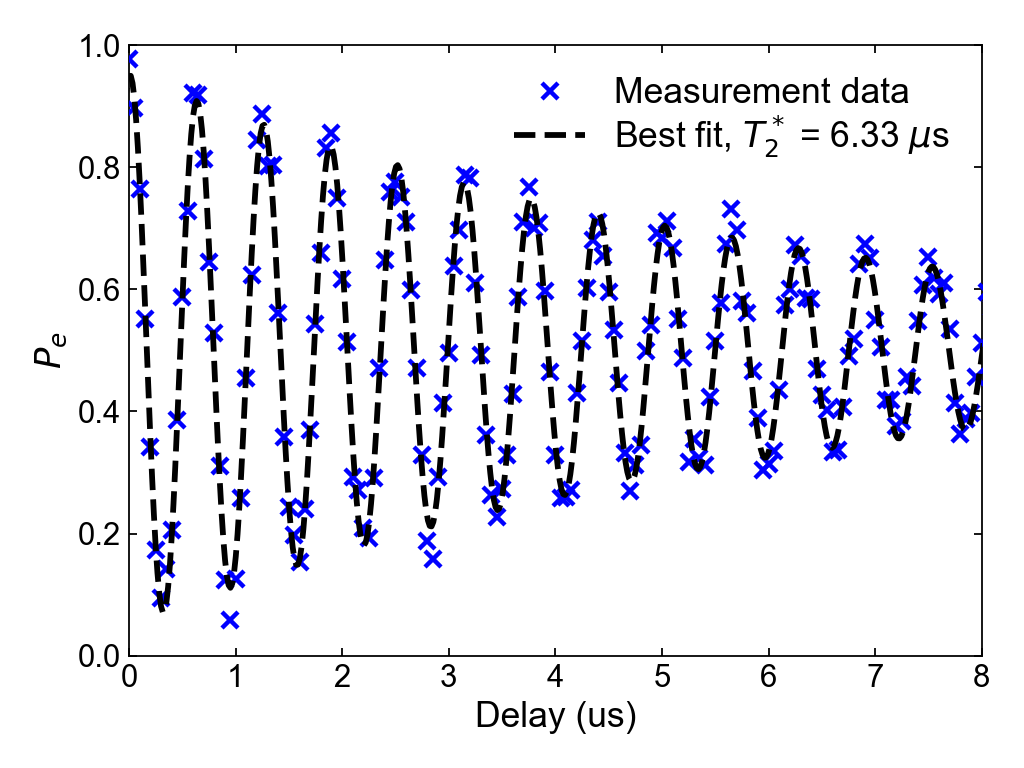}};
            \node[anchor=north west, text=black, font=\bfseries, xshift=-6pt, yshift=-6pt, overlay] at (image.north west) {d};
        \end{tikzpicture}
        \captionsetup{labelformat=empty, skip=0pt}%
        \caption{}%
        \label{fig:al_t2_star}
    \end{subfigure}
    \caption{\textbf{\boldmath Characterisation of the Al transmon at $50\,\mathrm{mK}$.} \textbf{a}~Punchout of the readout resonator, showing a shift from approximately $7.2885\,\mathrm{GHz}$ at low probe power to $7.2870\,\mathrm{GHz}$ after qubit saturation; \textbf{b}~Two-tone spectroscopy, with the $|0\rangle\rightarrow|1\rangle$ transition near $4.215\,\mathrm{GHz}$, the two-photon $|0\rangle\rightarrow|2\rangle$ transition near $4.10\,\mathrm{GHz}$, and the $|1\rangle\rightarrow|2\rangle$ transition near $3.99\,\mathrm{GHz}$; \textbf{c}~Energy relaxation, yielding $T_1=14.8\,\mu\mathrm{s}$; and \textbf{d}~Dephasing, yielding $T_2^*=6.3\,\mu\mathrm{s}$.}%
    \label{fig:al_characterisation}
\end{figure}

The Al readout resonator exhibits a punchout frequency shift of approximately $1.5\,\mathrm{MHz}$, as shown in Fig.~\ref{fig:al_punchout}. Two-tone spectroscopy in Fig.~\ref{fig:al_two_tone} resolves the $|0\rangle\rightarrow|1\rangle$ transition near $4.215\,\mathrm{GHz}$, the two-photon $|0\rangle\rightarrow|2\rangle$ transition near $4.10\,\mathrm{GHz}$, and the $|1\rangle\rightarrow|2\rangle$ transition near $3.99\,\mathrm{GHz}$. The time-domain measurements yield $T_1=14.8\,\mu\mathrm{s}$ and $T_2^*=6.3\,\mu\mathrm{s}$, as shown in Fig.~\ref{fig:al_t1} and Fig.~\ref{fig:al_t2_star}.

\bibliography{references}

\end{document}